\documentclass[11pt]{article}

\usepackage[letterpaper, margin=1in]{geometry}
\usepackage{amsmath,amssymb,amsfonts}
\usepackage{graphicx}
\usepackage{physics}
\usepackage{booktabs}
\usepackage{enumerate}
\usepackage{authblk}
\usepackage{xcolor}
\definecolor{darkgreen}{rgb}{0.0,0.5,0.0}
\usepackage[colorlinks,linkcolor=red,citecolor=darkgreen,urlcolor=blue]{hyperref}
\usepackage[capitalise]{cleveref}
\usepackage[amsmath,thmmarks]{ntheorem}

\theoremstyle{plain}
\theoremheaderfont{\normalfont\bfseries}
\theorembodyfont{\normalfont\itshape}
\theoremseparator{.}
\theoremsymbol{}
\newtheorem{theorem}{Theorem}
\newtheorem{lemma}[theorem]{Lemma}

\newtheorem{assumption}{Assumption}

\theoremstyle{nonumberplain}
\theoremheaderfont{\normalfont\itshape}
\theorembodyfont{\normalfont}
\theoremsymbol{\ensuremath{\square}}
\newtheorem{proof}{Proof}

\theoremstyle{nonumberplain}
\theoremheaderfont{\normalfont\bfseries}
\theorembodyfont{\normalfont}
\theoremsymbol{}

\renewcommand{\d}{\mathrm{d}}
\newcommand{\Ge}{\geqslant}
\newcommand{\Le}{\leqslant}
\newcommand{\T}{\top}
\newcommand{\avg}[1]{\langle{#1}\rangle}
\newcommand{\odd}{{\mathrm{odd}}}
\newcommand{\even}{{\mathrm{even}}}

\title{Optimal Convergence Rate of Lie--Trotter Approximation for Quantum Thermal Averages}
\author[1]{Xuda Ye\thanks{\texttt{ye481@purdue.edu}}}
\author[2]{Zhennan Zhou\thanks{\texttt{zhouzhennan@westlake.edu.cn}}}
\affil[1]{Department of Mathematics, Purdue University, West Lafayette, IN 47907, USA}
\affil[2]{Institute for Theoretical Sciences, Westlake University, Hangzhou, 310030, China}
\date{\today}

\hypersetup{
  pdftitle={Optimal Convergence Rate of Lie--Trotter Approximation for Quantum Thermal Averages},
  pdfauthor={Xuda Ye, Zhennan Zhou}
}

\begin{document}

\maketitle

\begin{abstract}
The Lie--Trotter product formula is a foundational approximation for the quantum partition function, yet obtaining rigorous error bounds for the unbounded Hamiltonians common in physics remains a challenge. This paper provides a quantitative error analysis for this approximation across two systems. For a particle in a smooth, periodic potential, we establish an optimal convergence rate of $\mathcal O(1/N^2)$ for both the partition function and thermal averages, where $N$ is the number of imaginary time steps. We then extend this analysis to the more challenging case of a confining potential on $\mathbb R$, proving a nearly optimal rate of $\mathcal O((\log N+1)^{1.5}/N^2)$. The derived error bounds provide a firm mathematical foundation for the second-order accuracy of path integral simulations in quantum statistical mechanics.


\end{abstract}

\section{Introduction}
Calculating the partition function is central to characterizing the canonical ensemble and its corresponding thermal averages at a given temperature \cite{davidson2013statistical,tuckerman2023statistical}. For quantum mechanical systems, however, this task becomes challenging \cite{feynman1966quantum,szabo1996modern,selsto2024computational}. The quantum partition function for a system described by the Hamiltonian operator $\hat H$ at an inverse temperature $\beta$ is given by
\begin{equation*}
	\mathcal Z = \tr\big[e^{-\beta\hat H}\big].
\end{equation*}
The computational cost of standard approaches, such as spectral methods or direct spatial discretization, grows exponentially with the system's dimension $d$---a well-known obstacle referred to as the curse of dimensionality \cite{tuckerman2023statistical,nielsen2010quantum}.

To circumvent this challenge, the path integral formulation introduced by R. Feynman provides a framework for approximating the partition function \cite{feynman1966quantum}. Consider a Hamiltonian composed of a kinetic term $\hat H_0$ and a potential term $\hat V$:
\begin{equation*}
	\hat H = \hat H_0 + \hat V = -\frac{1}{2m} \Delta + V(\hat x),
\end{equation*}
where $m>0$ is the particle mass, $\Delta$ is the Laplacian operator, and $V(\hat x)$ is the potential function. A common strategy is to apply the operator splitting, such as the Lie--Trotter product formula \cite{suzuki1976generalized,simon2005functional}, which yields an approximation for $\mathcal Z$:
\begin{equation*}
	\mathcal Z \approx \mathcal Z_N := \tr\Big[
	\big(e^{-\frac{\beta}{N}\hat H_0}
	e^{-\frac{\beta}{N}\hat V}\big)^N
	\Big].
\end{equation*}
By the cyclic symmetry of the trace, we can equivalently write
$$
\mathcal Z_N = \tr\Big[
\big(e^{-\frac{\beta}{2N}\hat V}e^{-\frac{\beta}{N}\hat H_0}
e^{-\frac{\beta}{2N}\hat V}\big)^N
\Big] = \tr\Big[
\big(e^{-\frac{\beta}{2N}\hat H_0}e^{-\frac{\beta}{N}\hat V}
e^{-\frac{\beta}{2N}\hat H_0}\big)^N
\Big].
$$
Furthermore, the thermal average of an observable $\hat O = O(\hat x)$, defined as
\begin{equation*}
	\avg{\hat O} = 
	\frac
	{\tr\big[e^{-\beta\hat H}\hat O\big]} 
	{\tr\big[e^{-\beta\hat H}\big]},
\end{equation*}
can be approximated using the same splitting:
\begin{equation*}
	\avg{\hat O}_N:=
	\frac{\tr\Big[
		\big(e^{-\frac{\beta}{N}\hat H_0}
		e^{-\frac{\beta}{N}\hat V}\big)^N \hat O
		\Big]}{
		\tr\Big[
		\big(e^{-\frac{\beta}{N}\hat H_0}
		e^{-\frac{\beta}{N}\hat V}\big)^N
		\Big]
	}.
\end{equation*}

This approximation scheme is the foundation of widely used computational methods, including path integral Monte Carlo (PIMC) \cite{ceperley1995path} and path integral molecular dynamics (PIMD) \cite{marx1996ab,ceriotti2010efficient,liu2016simple,ye2021efficient}, which enable the efficient calculation of quantum thermal averages. A key advantage is that the computational cost scales favorably with the system's dimension $d$ and the number of path integral replicas $N$, and the approximation becomes exact in the limit $N\to\infty$. 
Therefore, in practical simulations where $N$ must be finite, the convergence rate of this Lie--Trotter approximation is critical for guaranteeing computational efficiency and accuracy. Specifically, we aim to provide rigorous upper bounds on the differences between the partition functions $|\mathcal Z_N - \mathcal Z|$ and the thermal averages $|\avg{\hat O} - \avg{\hat O}_N|$.

Matrix analysis provides an alternate viewpoint on this problem. The Golden--Thompson inequality \cite{aujla2003weak,tao2010golden,forrester2014golden}, for any two Hermitian matrices $A$ and $B$, states that:
\begin{equation*}
	\tr\Big[\big(e^{\frac{A}{N}}e^{\frac{B}{N}}\big)^N\Big] \Ge \tr\big[e^{A+B}\big].
\end{equation*}
Applying this inequality to the quantum Hamiltonian by setting $A = -\beta \hat H_0$ and $B = -\beta \hat V$, we find that the approximate partition function is always an upper bound to the exact one:
\begin{equation}
	\mathcal Z_N = \tr\Big[
	\big(e^{-\frac{\beta}{N}\hat H_0}
	e^{-\frac{\beta}{N}\hat V}\big)^N
	\Big] \Ge \tr\big[e^{-\beta \hat H}\big] = \mathcal Z.
	\label{Golden}
\end{equation}
As a consequence, quantifying the error reduces to estimating the upper bound of $\mathcal Z_N - \mathcal Z$. A formal analysis of the symmetric operator splitting yields
\begin{equation*}
	\big(e^{-\frac{\beta}{2N} \hat V} e^{-\frac{\beta}{N}\hat H_0} e^{-\frac{\beta}{2N} \hat V} \big)^N = e^{-\beta \hat H} + \mathcal O(1/N^2),
\end{equation*}
which suggests a convergence rate for the partition function of $\mathcal O(1/N^2)$.
However, obtaining a mathematically rigorous estimate is challenging because the Hamiltonian $\hat H = \hat H_0 + \hat V$ is an unbounded operator on an infinite-dimensional space, making the analysis of the trace inherently difficult.

In this paper, we overcome the challenges posed by unbounded operators and provide a rigorous error analysis for the Lie--Trotter path integral approximation. For a quantum particle in a smooth, periodic potential, we establish an optimal convergence rate of $\mathcal O(1/N^2)$ for both the partition function and thermal averages:
\begin{equation*}
	0\Le \mathcal Z_N - \mathcal Z \Le \frac{C}{N^2} \qquad \text{and} \qquad \big|\avg{\hat O}_N - \avg{\hat O}\big| \Le \frac{C}{N^2}.
\end{equation*}
The constant $C>0$ is independent of $N$ and depends on physical parameters, including the mass $m$, inverse temperature $\beta$, and properties of the potential and observable. In particular, for the bound on $\mathcal Z_N - \mathcal Z$, we give the constant $C$ explicitly in closed form in \Cref{theorem: periodic,theorem: confining}. Furthermore, we extend our analysis to the more challenging case of a confining potential on $\mathbb R$. For this system, we establish a nearly optimal convergence rate:
\begin{equation*}
	0\Le \mathcal Z_N - \mathcal Z \Le \frac{C(\log N+1)^{\frac32}}{N^2} \qquad \text{and} \qquad \big|\avg{\hat O}_N - \avg{\hat O}\big| \Le \frac{C(\log N+1)^{\frac32}}{N^2}.
\end{equation*}

These results rigorously confirm that the Lie--Trotter product formula achieves second-order convergence for both periodic and confining potentials. They provide a firm theoretical foundation for the accuracy of widely used simulation methods such as PIMC and PIMD. A simple one-dimensional numerical test verifies the convergence order in the confining potential case (see \Cref{section: numerical}).

The remainder of this paper is organized as follows: Section~\ref{section: related} reviews related works; Section~\ref{section: state} presents the main convergence results; Section~\ref{section: path integral} outlines the path integral framework for the error analysis; Section~\ref{section: proof} contains the proofs of the main theorems; and Section~\ref{section: numerical} reports a one-dimensional numerical verification.   

\section{Related works}
\label{section: related}
The convergence of the Lie--Trotter product formula is a long-standing problem in mathematical physics. While strong convergence was established early on \cite{simon2005functional,tuckerman2023statistical}, obtaining quantitative error bounds is challenging, particularly for the unbounded Hamiltonians found in quantum mechanics. This is because properties of the operators, such as their domains, must be carefully handled to avoid divergent results \cite{burgarth2024strong}. Furthermore, much of the existing analysis has focused on bounding the operator norm of the error, which may not directly translate to the accuracy of thermal averages. Our work distinguishes itself by directly analyzing the trace error, a quantity of more direct physical relevance for statistical mechanics.

The error of the Lie--Trotter product formula is quantified by the operator
\begin{equation*}
	\mathcal D_N :=
	\big(e^{-\frac{\beta}{2N}\hat V}
	e^{-\frac{\beta}{N}\hat H_0}
	e^{-\frac{\beta}{2N}\hat V}\big)^N - 
	e^{-\beta\hat H},
\end{equation*}
where the physically relevant quantity for the partition function is the trace error, $\tr[\mathcal D_N] = \mathcal Z_N - \mathcal Z$. For a periodic potential $V(x)$ on a finite interval $[0,L]$, the trace error is related to the operator norm via the inequality:
\begin{equation*}
	0\Le \tr[\mathcal D_N] \Le L \norm{\mathcal D_N}.
\end{equation*}
However, this straightforward relationship does not hold for confining potentials on the infinite domain $\mathbb R$, making the trace error analysis more complex.

Early works using functional analysis established an operator norm bound of $\mathcal O(\log N/N)$ \cite{rogava1993error,neidhardt1998error}. This rate was later improved to $\mathcal O(1/N)$ using both functional analysis \cite{ichinose2001norm,ichinose2001note} and path integral techniques \cite{takanobu1997error,ichinose1997estimate,doumeki1998error}. A recent paper also provides an operator norm bound for potential functions with low regularity \cite{becker2025convergence}.
While formal considerations suggest an $\mathcal O(1/N^2)$ rate should be attainable---and Suzuki's hierarchy of higher-order product formulas yields $\mathcal O(1/N^{2j})$ at the $(2j)$-th order \cite{suzuki1991general,hatano2005finding}---rigorously proving even the second-order rate for unbounded operators has remained challenging. Our work establishes the optimal $\mathcal O(1/N^2)$ convergence rate for the trace error with periodic potentials, and a nearly optimal $\mathcal O((\log N+1)^{\frac32}/N^2)$ rate for confining potentials. Table~\ref{table: comparison} summarizes these results and contrasts our focus on trace error with prior work on norm error.

\begin{table}[htbp]
	\footnotesize
	\caption{Comparison of error analysis results for the Lie--Trotter approximation. The exponent $\delta$ in the low-regularity row depends on the regularity of the potential function $V(x)$.}
	\label{table: comparison}
	\begin{center}
	\begin{tabular}{ccccc}
		\toprule
		\textbf{Reference(s)} & \textbf{Potential} & \textbf{Error} & \textbf{Result} & \textbf{Proof} \vspace{4pt} \\
		\midrule
		\cite{rogava1993error,neidhardt1998error} &
		regular & norm & $\norm{\mathcal D_N} \Le \dfrac{C\log N}{N}$ & functional analysis \vspace{4pt}\\
		\cite{takanobu1997error,ichinose1997estimate,doumeki1998error} & regular & norm & $\norm{\mathcal D_N} \Le \dfrac{C}{N}$ & path integral \vspace{4pt} \\
		\cite{ichinose2001norm,ichinose2001note} & regular & norm & $\norm{\mathcal D_N} \Le \dfrac{C}{N}$ & functional analysis \vspace{4pt} \\
		\cite{becker2025convergence} & low-regular. & norm & $\norm{\mathcal D_N} \Le \dfrac{C}{N^\delta}$ & functional analysis \vspace{4pt} \\
		Theorem~\ref{theorem: periodic} & periodic & trace & $ \tr[\mathcal D_N] \Le \dfrac{C}{N^2}$ & path integral \vspace{4pt} \\
		Theorem~\ref{theorem: confining} & confining & trace & $ \tr[\mathcal D_N] \Le \dfrac{C(\log N+1)^{\frac32}}{N^2}$ & path integral \vspace{4pt} \\
		\bottomrule
	\end{tabular}
	\end{center}
\end{table}

The challenge of establishing an optimal rate for unbounded operators is mirrored in studies of the real-time Schr\"odinger equation. For the real-time evolution, the corresponding error operator for a symmetric splitting is given by
\begin{equation*}
	\mathcal E_N := \big(e^{-\frac{it}{2N}\hat A}
	e^{-\frac{it}{N}\hat B}
	e^{-\frac{it}{2N}\hat A}\big)^N - 
	e^{-it(\hat A+\hat B)},
\end{equation*}
where $\hat A$ and $\hat B$ are Hermitian operators. The error analysis in this setting is notoriously difficult due to the dispersive nature of the Schr\"odinger equation. For the case where $\hat A$ and $\hat B$ are bounded operators, an optimal $\mathcal O(1/N^2)$ error bound on the operator norm has been established \cite{childs2021theory}:
\begin{equation*}
	\norm{\mathcal E_N} \Le \frac{t^3}{12N^2} 
	\big\|[\hat B,[\hat B,\hat A]]\big\| + 
	\frac{t^3}{24N^2} \big\|[\hat A,[\hat A,\hat B]]\big\|,
\end{equation*}
where $[\hat A,\hat B] = \hat A\hat B-\hat B\hat A$ is the commutator. For the unbounded operators common in quantum mechanics, sharp local-error representations and norm bounds for Schr\"odinger splittings have been developed in the numerical-analysis literature \cite{jahnke2000error,descombes2010exact,lubich2008quantum}, yet achieving the optimal rate under physically relevant operator-domain conditions remains an active topic \cite{burgarth2024strong,becker2025convergence}.

\section{Statement of main results}
\label{section: state}
In this section, we state the main convergence results for periodic and confining potentials. The Hamiltonian operator $\hat H$ of the system is given by
$$
\hat H = \hat H_0 + \hat V = -\frac{1}{2m}\Delta + V(\hat x).
$$
The exact partition function $\mathcal Z$ and its Lie--Trotter approximation $\mathcal Z_N$ are
\begin{equation*}
	\mathcal Z = \tr\big[e^{-\beta\hat H}\big], \qquad 
	\mathcal Z_N = \tr\Big[\big(e^{-\frac{\beta}{N}\hat H_0} e^{-\frac{\beta}{N}\hat V}\big)^N\Big],
\end{equation*}
while the trace quantities corresponding to the observable operator $\hat O$ are
\begin{equation*}
	\mathcal W = \tr\big[e^{-\beta\hat H} \hat O\big], \qquad 
	\mathcal W_N = \tr\Big[\big(e^{-\frac{\beta}{N}\hat H_0} e^{-\frac{\beta}{N}\hat V}\big)^N \hat O\Big].
\end{equation*}
Our goal is to quantify the Lie--Trotter approximation error $\mathcal Z_N - \mathcal Z$ and $\mathcal W_N - \mathcal W$.

Our first main result concerns quantum systems with a smooth, periodic potential. We impose the following conditions on the potential $V(x)$ and the observable $O(x)$.
\begin{assumption}[periodic]
	\label{assumption: periodic}
	The potential $V(x)$ and the observable $O(x)$ are periodic and $C^2$ on $[0,L]$, with $V(x) \Ge 0$. For some constant $K>0$,
	\begin{equation*}
		\sup_{x\in [0,L]} \Big(1+|V(x)| + |V'(x)| + |V''(x)| + |O(x)| + |O'(x)| + |O''(x)| \Big) \Le K.
	\end{equation*}
\end{assumption}
We establish an optimal convergence rate for this Lie--Trotter approximation.
\begin{theorem}[periodic]
	\label{theorem: periodic}
	Under Assumption~\ref{assumption: periodic}, $\mathcal Z_N \Ge \mathcal Z$, and
	\begin{equation*}
		|\mathcal Z_N - \mathcal Z|,\ |\mathcal W_N - \mathcal W| \Le \frac{67 K^4 L \beta(1+\beta^2)}{m N^2}\bigg(\frac{\beta}{m}+1\bigg).
	\end{equation*}
\end{theorem}

Next, we extend the analysis to the more challenging case of a confining potential on the entire real line, $\mathbb R$. The required assumptions are as follows.
\begin{assumption}[confining]
	\label{assumption: confining}
	The potential $V(x)$ and the observable $O(x)$ are $C^2$ on $\mathbb R$. For some constants $a, K>0$, $V(x)$ has at least quadratic growth
	\begin{equation*}
		V(x) \Ge \frac{a}{2}x^2, \qquad \forall x\in\mathbb R,
	\end{equation*}
	and
	\begin{equation*}
		\sup_{x\in\mathbb R} \bigg(1+|V(0)| + \frac{|V'(x)|}{|x|+1} + |V''(x)| + |O(x)| + |O'(x)| + |O''(x)|\bigg) \Le K.
	\end{equation*}
\end{assumption}
We establish a nearly optimal rate with a logarithmic factor.
\begin{theorem}[confining]
	\label{theorem: confining}
	Under Assumption~\ref{assumption: confining}, $\mathcal Z_N \Ge \mathcal Z$, and
	\begin{equation*}
		|\mathcal Z_N - \mathcal Z|,\ |\mathcal W_N - \mathcal W| \Le \bigg(\frac{3840\sqrt 2\, K^4 \beta(1+\beta^2)(\beta/m+1)}{m\, c^{\frac32}} + 9 C_R\bigg)\frac{(\log N+1)^{\frac32}}{N^2},
	\end{equation*}
	where $c = \dfrac{a\beta}{2(1+a\beta^2/(4m))}$ and $C_R = \dfrac{K}{c}\sqrt{\dfrac{m}{2\pi\beta}}$.
\end{theorem}

The bounds on $|\mathcal Z_N - \mathcal Z|$ and $|\mathcal W_N - \mathcal W|$ in Theorems~\ref{theorem: periodic} and~\ref{theorem: confining} also yield a bound on the thermal average $\avg{\hat O} = \mathcal W/\mathcal Z$. By the Golden--Thompson inequality, $\mathcal Z_N \Ge \mathcal Z > 0$ for every $N$, so $1/\mathcal Z_N \Le 1/\mathcal Z$ and the sequences $\mathcal Z_N, \mathcal W_N$ are uniformly bounded. The identity
\begin{equation*}
	\frac{\mathcal W_N}{\mathcal Z_N} - \frac{\mathcal W}{\mathcal Z} = \frac{(\mathcal W_N - \mathcal W)\mathcal Z - \mathcal W(\mathcal Z_N - \mathcal Z)}{\mathcal Z_N \mathcal Z}
\end{equation*}
combined with $1/\mathcal Z_N \Le 1/\mathcal Z$ gives
\begin{equation*}
	\big|\avg{\hat O}_N - \avg{\hat O}\big| = \bigg|\frac{\mathcal W_N}{\mathcal Z_N} - \frac{\mathcal W}{\mathcal Z}\bigg| \Le \frac{|\mathcal W_N - \mathcal W|}{\mathcal Z} + \frac{|\mathcal W|\,|\mathcal Z_N - \mathcal Z|}{\mathcal Z^2}.
\end{equation*}
Substituting the bounds from Theorems~\ref{theorem: periodic} and \ref{theorem: confining},
\begin{equation*}
	\big|\avg{\hat O}_N - \avg{\hat O}\big|
	\Le \left\{
	\begin{aligned}
		& \frac{C_{\beta,m}}{N^2}, && \text{for periodic potential}, \\
		& \frac{C_{\beta,m}(\log N+1)^{\frac32}}{N^2}, &&
		\text{for confining potential},
	\end{aligned}
	\right.
\end{equation*}
where $C_{\beta,m}>0$ depends on $\beta$, $m$, and the assumption constants, but is independent of $N$; we do not track its explicit form.
The factor $(\log N+1)^{\frac32}$ in Theorem~\ref{theorem: confining} is primarily an artifact of the proof technique: the domain truncation $[-R,R]$ used to control the unbounded integration domain forces a balance $R^2 \sim \log N$ between the algebraic and exponential error terms, which propagates into the final bound. An alternative operator splitting that bypasses domain truncation recovers the optimal $\mathcal O(1/N^2)$ rate; see Appendix~\ref{appendix: alternate}.

\section{Path integral framework for error analysis}
\label{section: path integral}

We now develop the path integral framework underpinning our error analysis. We derive path integral representations of the partition function $\mathcal Z = \tr\big[e^{-\beta\hat H}\big]$, the observable trace $\mathcal W = \tr\big[e^{-\beta\hat H}\hat O\big]$, and their Lie--Trotter approximations $\mathcal Z_N$ and $\mathcal W_N$. Within this representation, the successive difference $\mathcal Z_N - \mathcal Z_{2N}$ takes the form of a non-negative path integral, which yields a path integral interpretation of the Araki--Lieb--Thirring inequality and is central to our error bound.

For convenience, throughout this section we write out the derivations only for the periodic case, in which $V(x)$ is periodic on $[0,L]$. The confining case proceeds in the same way: every formula carries over once the base-point integral $\int_0^L \d x_0$ is replaced by $\int_{\mathbb R} \d x_0$.

\subsection{Representation of trace quantities}
Throughout this section we adopt the bra-ket notation in quantum mechanics \cite{nielsen2010quantum}; the element $\mel{x}{\hat A}{y}$ may be read as the integral kernel (Green's function) of the operator $\hat A$. Inserting a complete set of position eigenstates $\ket{x_0}$ then expresses the trace quantity $\mathcal W$ as an integral over the diagonal of the propagator in the position basis:
\begin{equation}
	\mathcal W = \tr\big[e^{-\beta \hat H} \hat O\big] = \int_0^L \mel{x_0}{e^{-\beta\hat H}}{x_0} O(x_0) \d x_0.
	\label{Z integral}
\end{equation}
Since $V(x)$ is periodic on $[0,L]$, the diagonal propagator $\mel{x_0}{e^{-\beta\hat H}}{x_0}$ is periodic in $x_0$, and integrating over a single period reproduces the full trace.

As an illustration, recall that $\hat H_0 = -\frac{1}{2m}\Delta$ is the free-particle Schr\"odinger operator; its propagator $\mel{x_0}{e^{-\beta\hat H_0}}{x_0}$ is constant in $x_0$, so \eqref{Z integral} reduces to
\begin{equation*}
	\tr\big[e^{-\beta \hat H_0} \hat O\big] =
	\int_0^L \mel{x_0}{e^{-\beta\hat H_0}}{x_0} O(x_0)\d x_0 =
	\sqrt{\frac{m}{2\pi\beta}}
	\int_0^L O(x_0) \d x_0.
\end{equation*}

For the Lie--Trotter approximation $\mathcal W_N$, repeated insertion of position eigenstates $\{\ket{x_j}\}_{j=1}^{N-1}$ discretizes the imaginary time propagation into $N$ steps:
\begingroup\allowdisplaybreaks
\begin{align*}
	& ~~~~\mathcal W_N = \tr\Big[
	\big(e^{-\frac{\beta}{N}\hat H_0}
	e^{-\frac{\beta}{N}\hat V}\big)^N \hat O
	\Big] \notag \\
	& = \int_0^L \mel{x_0}{\big(e^{-\frac{\beta}{N}\hat H_0}
		e^{-\frac{\beta}{N}\hat V}\big)^N}{x_0} O(x_0) \d x_0 \notag \\
	& = \int_0^L \d x_0 \int_{\mathbb R^{N-1}} O(x_0)
	\prod_{j=0}^{N-1} \mel{x_{j+1}}
	{e^{-\frac{\beta}{N}\hat H_0} e^{-\frac{\beta}{N}\hat V}}
	{x_j}
	\d x_1 \cdots \d x_{N-1} \notag \\
	& = \int_0^L \d x_0 \int_{\mathbb R^{N-1}} O(x_0)
	\prod_{j=0}^{N-1} e^{-\frac{\beta}{N}V(x_j)} \mel{x_{j+1}}
	{e^{-\frac{\beta}{N}\hat H_0}}
	{x_j}
	\d x_1 \cdots \d x_{N-1} \notag \\
	& =
	\int_0^L \d x_0 \times \bigg(\frac{mN}{2\pi\beta}\bigg)^{\frac N2}\int_{\mathbb R^{N-1}} O(x_0) \times \cdots\\
	& \hspace{1.8cm} 
	\exp\bigg(-\frac{\beta}{N} \sum_{j=0}^{N-1} V(x_j) - \frac{mN}{2\beta} \sum_{j=0}^{N-1} (x_{j+1}-x_j)^2\bigg)
	\d x_1 \cdots \d x_{N-1},
\end{align*}
\endgroup
where we identify $x_N \equiv x_0$ to close the path under the trace. The third line uses the fact that $\hat V$ is diagonal in the position basis, $\mel{x'}{\hat V}{x} = V(x)\delta(x'-x)$, and the last line uses the free-particle propagator. The exponent defines an effective classical potential on the path variables $\{x_j\}_{j=0}^{N-1}$:
\begin{equation}
	U(x_0,x_1,\cdots,x_{N-1}) =
	\frac{\beta}{N} \sum_{j=0}^{N-1} V(x_j) + \frac{mN}{2\beta} \sum_{j=0}^{N-1} (x_{j+1}-x_j)^2,
\end{equation}
which is precisely the ring polymer potential underlying path integral molecular dynamics \cite{liu2016simple,ye2021efficient}. In terms of $U$, the observable trace $\mathcal W_N$ becomes
\begin{equation}
	\mathcal W_N = \bigg(\frac{mN}{2\pi\beta}\bigg)^{\frac N2} \int_0^L \d x_0 \int_{\mathbb R^{N-1}}
	O(x_0) \times\exp\big(-U(x_0,x_1,\cdots,x_{N-1})\big) \d x_1 \cdots \d x_{N-1}.
	\label{eq: W_N U}
\end{equation}

To recast the expression \eqref{eq: W_N U} probabilistically, we define a conditional probability density for the path variables $\{x_1, \dots, x_{N-1}\}$ given the starting point $x_0$:
\begin{align}
	\sigma_N[x_0](x_1,\cdots,x_{N-1}) & :=
	\frac{1}{\mel{x_0}{e^{-\beta \hat H_0}}{x_0}}{\prod_{j=0}^{N-1} \mel{x_{j+1}}{e^{-\frac{\beta}{N} \hat H_0}}{x_j}} \notag \\
	& = \sqrt{\frac{2\pi\beta}{m}}\bigg(\frac{mN}{2\pi\beta}\bigg)^{\frac{N}{2}}
	\exp\bigg(- \frac{mN}{2\beta} \sum_{j=0}^{N-1} (x_{j+1}-x_j)^2\bigg),
	\label{sigma_N definition}
\end{align}
where the closure $x_N \equiv x_0$ is implicit in the product. The bracketed argument $[x_0]$ denotes the conditioning on $x_0$, and $\sigma_N[x_0]$ is a probability density on $\mathbb R^{N-1}$.

The distribution $\sigma_N[x_0]$ is the law of a discrete Brownian bridge pinned at $x_0$, and its key properties are summarized below (in the lemma we assume $x_0 = 0$).
\begin{lemma}
	\label{lemma: Brownian}
	The random variables $x_1,\cdots,x_{N-1}$ from the distribution $\sigma_N[0]$ have the same joint distribution as the process
	$$
	\sqrt{\frac{\beta}{m}}\bigg(B_{j/N} - \frac{j}{N} B_1\bigg),\qquad j =1,\cdots,N-1,
	$$
	where $B_t$ is the standard one-dimensional Brownian motion. Furthermore,
	\begin{equation*}
		\mathbb E\big[x_j^2\big] = \frac{\beta}{m} \frac{j(N-j)}{N^2},\qquad
		\mathbb E\big[(x_{j+1}-x_j)^2\big] = \frac{\beta}{m}\frac{N-1}{N^2}.
	\end{equation*}
	And the conditional expectations satisfy
	\begin{align*}
		\mathbb E\big[x_{j-1}\big|x_j\big] & = 
		\frac{j-1}{j} x_j, & \quad 
		\mathbb E\big[(x_{j-1}-x_j)^2\big|x_j\big] & = 
		\frac{\beta}{m} \frac{j-1}{jN} + \frac{x_j^2}{j^2}, \\
		\mathbb E\big[x_{j+1}\big|x_j\big] & = 
		\frac{N-j-1}{N-j}x_j, &\quad
		\mathbb E\big[(x_{j+1}-x_j)^2\big|x_j\big] &=
		\frac{\beta}{m} \frac{N-j-1}{N(N-j)} + \frac{x_j^2}{(N-j)^2}.
	\end{align*}
\end{lemma}
\noindent The proof of \Cref{lemma: Brownian} is a standard stochastic calculus computation and is deferred to Appendix~\ref{appendix: lemma proof}.

Using the density $\sigma_N[x_0]$, we may rewrite $\mathcal W_N$ as an expectation:
\begin{equation*}
	\mathcal W_N = \sqrt{\frac{m}{2\pi\beta}}\int_0^L \d x_0
	\times \mathbb E_{\sigma_N[x_0]}
	\bigg[ O(x_0) \times
	\exp\bigg(-\frac{\beta}{N}
	\sum_{j=0}^{N-1} V(x_j)\bigg)
	\bigg],
\end{equation*}
where $\mathbb E_{\sigma_N[x_0]}$ denotes the expectation over $x_1,\cdots,x_{N-1}$ drawn from $\sigma_N[x_0]$. The cyclic symmetry of the path integral under permutation of the time slices allows us to replace $O(x_0)$ by its average over any subset of the path variables. Averaging over all $N$ beads yields
\begin{equation}
	\mathcal W_N = \sqrt{\frac{m}{2\pi\beta}}\int_0^L \d x_0
	\times \mathbb E_{\sigma_N[x_0]}
	\bigg[\frac1N \sum_{j=0}^{N-1} O(x_j)\times
	\exp\bigg(-\frac{\beta}{N}
	\sum_{j=0}^{N-1} V(x_j)\bigg)
	\bigg].
	\label{trace N PI}
\end{equation}
Equation \eqref{trace N PI} provides a probabilistic interpretation of the Lie--Trotter approximation: $\mathcal W_N$ is the average of the observable along free-particle bridges, weighted by the discrete Boltzmann factor of the potential.

In the continuum limit $N\to\infty$, the discrete path distribution $\sigma_N[x_0]$ converges to the Gaussian measure $\sigma[x_0]$ of a continuous Brownian bridge $\{x(\tau)\}_{0\Le\tau\Le 1}$ pinned at $x(0)=x(1)=x_0$ \cite{lu2020continuum,ye2023dimension}, which admits the explicit representation
\begin{equation}
	x(\tau) = x_0 + \sqrt{\frac{\beta}{m}}\big(B_{\tau} - \tau B_1\big),\qquad
	\tau \in [0,1].
	\label{continuous Brownian}
\end{equation}
In this limit, the Riemann sums in \eqref{trace N PI} pass to their continuous counterparts,
\begin{equation*}
	\frac1N \sum_{j=0}^{N-1} O(x_j) \to \int_0^1 O(x(\tau)) \d\tau, \qquad
	\frac1N \sum_{j=0}^{N-1} V(x_j) \to \int_0^1 V(x(\tau)) \d\tau,
\end{equation*}
and \eqref{trace N PI} formally yields
\begin{equation}
	\lim_{N\to\infty}
	\mathcal W_N = \sqrt{\frac{m}{2\pi\beta}}\int_0^L \d x_0
	\times \mathbb E_{\sigma[x_0]}
	\bigg[\int_0^1 O(x(\tau))\d\tau \times
	\exp\bigg(-\beta\int_0^1 V(x(\tau))\d\tau\bigg)
	\bigg].
	\label{trace PI}
\end{equation}
The right-hand side of \eqref{trace PI} is precisely the Feynman--Kac representation of $\mathcal W = \tr\big[e^{-\beta\hat H}\hat O\big]$ \cite{simon2005functional}, so $\lim_{N\to\infty}\mathcal W_N = \mathcal W$. The following lemma makes this convergence quantitative under either of our standing assumptions.
\begin{lemma}
	\label{lemma: consistency}
	Under either Assumption~\ref{assumption: periodic} or \ref{assumption: confining}, the trace quantities $\mathcal W = \tr\big[e^{-\beta\hat H} \hat O\big]$ and $\mathcal W_N = \tr\Big[
	\big(e^{-\frac{\beta}{N}\hat H_0}
	e^{-\frac{\beta}{N}\hat V}\big)^N \hat O
	\Big]$ satisfy $\lim_{N\to\infty} \mathcal W_N = \mathcal W$. To be specific,
	\begin{equation*}
		|\mathcal W_N - \mathcal W| \Le 
		\left\{
		\begin{aligned}
			& \frac{C_{\beta,m}}{\sqrt{N}}, && \text{for periodic potential}, \\
			& \frac{C_{\beta,m}(\log N+1)}{\sqrt{N}}, &&
			\text{for confining potential},
		\end{aligned}
		\right.
	\end{equation*}
	where $C_{\beta,m}>0$ depends on $\beta$ and $m$, but not on $N$.
\end{lemma}
The proof of \Cref{lemma: consistency} is provided in Appendix~\ref{appendix: consistency}. \Cref{lemma: consistency} only establishes consistency of the Lie--Trotter approximation: the rate obtained here is $\mathcal O(1/\sqrt{N})$, well below the $\mathcal O(1/N^2)$ rate established in our main theorems, and its proof uses only standard Brownian-bridge estimates. The analysis leading to the optimal second-order rate $\mathcal O(1/N^2)$ is the subject of the main theorems stated in \Cref{section: state}.

\subsection{Araki--Lieb--Thirring inequality}

A key result from matrix theory is the Araki--Lieb--Thirring inequality \cite{araki1990inequality,audenaert2007araki,hiai2014introduction}, which states that for any positive semidefinite matrices $A$ and $B$,
\begin{equation}
	\tr\big[
	(A^{\frac t2}B^t A^{\frac t2})^{s}
	\big]\Ge \tr\big[
	(A^{\frac12}BA^{\frac12})^{st}
	\big],\qquad \forall s>0, ~~ t\Ge 1.
	\label{ALT}
\end{equation}
By setting $s=N$ and $t=2$ in \eqref{ALT}, we obtain the specific case
\begin{equation}
	\tr\big[(AB^2A)^N\big] \Ge \tr\big[(A^{\frac12}BA^{\frac12})^{2N}\big].
	\label{ALT 2}
\end{equation}
Although the inequality \eqref{ALT 2} only holds for finite-dimensional matrices, we may formally choose $A = e^{-\frac{\beta}{2N}\hat H_0}$ and $B = e^{-\frac{\beta}{2N}\hat V}$ for the Hamiltonian operator $\hat H = \hat H_0 + \hat V$. Using the identity $\tr[(XY)^N] = \tr\big[(X^{\frac12}YX^{\frac12})^N\big]$, the inequality \eqref{ALT 2} becomes
\begin{equation}
	\mathcal Z_N = \tr\Big[
	\big(e^{-\frac{\beta}{N}\hat H_0}
	e^{-\frac{\beta}{N}\hat V}\big)^N
	\Big] \Ge \mathcal Z_{2N} =
	\tr\Big[
	\big(e^{-\frac{\beta}{2N}\hat H_0}
	e^{-\frac{\beta}{2N}\hat V}\big)^{2N}
	\Big],
	\label{ALT N}
\end{equation}
which is simply $\mathcal Z_N \Ge \mathcal Z_{2N}$. This demonstrates that $\{\mathcal Z_{2^j N}\}_{j=0}^\infty$ is monotonically decreasing. Furthermore, this sequence converges to the exact partition function,
\begin{equation*}
	\mathcal Z_N \Ge \lim_{j\to\infty}
	\mathcal Z_{2^j N} = \mathcal Z =
	\tr\big[e^{-\beta\hat H}\big],
\end{equation*}
and we recover the Golden--Thompson inequality \eqref{Golden}. This motivates us to estimate the successive difference $\mathcal W_N - \mathcal W_{2N}$ rather than the total difference $\mathcal W_N - \mathcal W$.

We now develop a path integral representation for the successive difference $\mathcal W_N - \mathcal W_{2N}$. We derive two equivalent expressions for $\mathcal W_N$ on a $2N$-step path, which we then compare directly with $\mathcal W_{2N}$. The first expression, the even-steps representation, arises from a symmetric grouping of operators. We insert a complete set of position eigenstates $\{\ket{x_j}\}_{j=0}^{2N-1}$ into the expression for $\mathcal W_N$:
\begin{align*}
	& ~~~~ \mathcal W_N = \tr\Big[
	\big(e^{-\frac{\beta}{2N}\hat V}e^{-\frac{\beta}{N}\hat H_0}
	e^{-\frac{\beta}{2N}\hat V}\big)^N \hat O
	\Big] \notag \\
	& = \int_0^L \mel{x_0}{\big(e^{-\frac{\beta}{2N}\hat V}e^{-\frac{\beta}{N}\hat H_0}
		e^{-\frac{\beta}{2N}\hat V}\big)^N}{x_0} O(x_0)\d x_0 = \int_0^L \d x_0 \int_{\mathbb R^{2N-1}}  O(x_0) \, \times \\
	& ~~~~
	\prod_{j=0}^{N-1} e^{-\frac{\beta}{N} V(x_{2j})} \mel{x_{2j+2}}{e^{-\frac{\beta}{2N} \hat H_0}}{x_{2j+1}}
	\mel{x_{2j+1}}{e^{-\frac{\beta}{2N} \hat H_0}}{x_{2j}} \d x_1\cdots \d x_{2N-1} \notag \\
	& = \sqrt{\frac{m}{2\pi\beta}}\int_0^L  \d x_0 
	\times \mathbb E_{\sigma_{2N}[x_0]}
	\bigg[O(x_0) \exp\bigg(-\frac{\beta}N \sum_{j=0}^{N-1} V(x_{2j})\bigg)\bigg].
\end{align*}
Due to the cyclic symmetry of the path integral under the trace, the observable can be averaged over all even time slices. This gives the even-steps representation:
\begin{equation}
	\mathcal W_N = \sqrt{\frac{m}{2\pi\beta}}\int_0^L 
	\d x_0 
	\times \mathbb E_{\sigma_{2N}[x_0]}
	\bigg[\frac1N \sum_{j=0}^{N-1} O(x_{2j})\times\exp\bigg(-\frac{\beta}N \sum_{j=0}^{N-1} V(x_{2j})\bigg)\bigg].
	\label{Z_N 1}
\end{equation}
An alternative, equivalent expression can be derived using a different operator splitting that places the potential on the odd time slices:
\begin{align*}
	& ~~~~ \mathcal W_N = \tr\Big[
	\big(e^{-\frac{\beta}{2N}\hat H_0}e^{-\frac{\beta}{N}\hat V}
	e^{-\frac{\beta}{2N}\hat H_0}\big)^{N-1}
	\big(e^{-\frac{\beta}{2N}\hat H_0}e^{-\frac{\beta}{2N}\hat V} \hat O
	e^{-\frac{\beta}{2N}\hat V} 
	e^{-\frac{\beta}{2N}\hat H_0}\big)
	\Big] \notag \\
	& = \int_0^L O(x_1) \times e^{-\frac \beta N V(x_1)} \mel{x_0}{\big(e^{-\frac{\beta}{2N}\hat H_0}e^{-\frac{\beta}{N}\hat V}
	e^{-\frac{\beta}{2N}\hat H_0}\big)^{N-1}}{x_1} \mel{x_1}{e^{-\frac{\beta}{2N}\hat H_0}}{x_0}\d x_0 \d x_1 \\
	& = \int_0^L \d x_0 \int_{\mathbb R^{2N-1}} O(x_1) \, \times 
	\prod_{j=0}^{N-1} e^{-\frac{\beta}{N} V(x_{2j+1})} \mel{x_{2j+2}}{e^{-\frac{\beta}{2N} \hat H_0}}{x_{2j+1}}
	\mel{x_{2j+1}}{e^{-\frac{\beta}{2N} \hat H_0}}{x_{2j}} \\
	& ~~~~ \d x_1\cdots \d x_{2N-1} = \sqrt{\frac{m}{2\pi\beta}}\int_0^L \d x_0 
	\times \mathbb E_{\sigma_{2N}[x_0]}
	\bigg[O(x_1) \exp\bigg(-\frac{\beta}N \sum_{j=0}^{N-1} V(x_{2j+1})\bigg)\bigg].
\end{align*}
Again invoking the cyclic symmetry to average the observable over all odd-indexed path variables, we obtain the odd-steps representation:
\begin{equation}
	\mathcal W_N = \sqrt{\frac{m}{2\pi\beta}}\int_0^L 
	\d x_0 
	\times \mathbb E_{\sigma_{2N}[x_0]}
	\bigg[\frac1N \sum_{j=0}^{N-1} O(x_{2j+1})\times \exp\bigg(-\frac{\beta}N \sum_{j=0}^{N-1} V(x_{2j+1})\bigg)\bigg].
	\label{Z_N 2}
\end{equation}

For notational convenience, we define the even and odd averages
\begin{gather*}
	\bar V_{N,\even} = \frac1N \sum_{j=0}^{N-1} V(x_{2j}),\qquad 
	\bar V_{N,\odd} = \frac1N \sum_{j=0}^{N-1} V(x_{2j+1}), \\
	\bar O_{N,\even} = \frac1N \sum_{j=0}^{N-1} O(x_{2j}),\qquad 
	\bar O_{N,\odd} = \frac1N \sum_{j=0}^{N-1} O(x_{2j+1}).
\end{gather*}
Then the odd- and even-steps representations of $\mathcal W_N$ in \eqref{Z_N 1} and \eqref{Z_N 2} are written as
\begin{align}
	\mathcal W_N & = \sqrt{\frac{m}{2\pi\beta}}\int_0^L 
	\d x_0 
	\times \mathbb E_{\sigma_{2N}[x_0]}
	\Big[\bar O_{N,\even}\times e^{-\beta \bar V_{N,\even}}\Big] \notag \\
	& = \sqrt{\frac{m}{2\pi\beta}}\int_0^L 
	\d x_0 
	\times \mathbb E_{\sigma_{2N}[x_0]}
	\Big[\bar O_{N,\odd}\times e^{-\beta \bar V_{N,\odd}}\Big].
	\label{Z_N 3}
\end{align}
For comparison, the path integral representation for $\mathcal W_{2N}$ is given directly by
\begin{equation}
	\mathcal W_{2N} = \sqrt{\frac{m}{2\pi\beta}}\int_0^L \d x_0 
	\times \mathbb E_{\sigma_{2N}[x_0]}
	\bigg[\frac12\big(\bar O_{N,\even} + \bar O_{N,\odd}\big) \times e^{-\frac{\beta}{2}\big(\bar V_{N,\even}+\bar V_{N,\odd}\big)}\bigg].
	\label{Z_N 4}
\end{equation}
From the equalities \eqref{Z_N 3} and \eqref{Z_N 4}, the difference $\mathcal W_N - \mathcal W_{2N}$ can be expressed as:
	\begin{align}
		& \mathcal W_N - \mathcal W_{2N} = \frac12 \sqrt{\frac{m}{2\pi\beta}}\int_0^L \d x_0 
		\,\times \notag \\
	 \mathbb E_{\sigma_{2N}[x_0]} & 
		\bigg[\Big(
		\bar O_{N,\even} e^{-\frac{\beta}2 \bar V_{N,\even}} - 
		\bar O_{N,\odd} e^{-\frac{\beta}2 \bar V_{N,\odd}}
		\Big)\Big( e^{-\frac{\beta}2 \bar V_{N,\even}} - e^{-\frac{\beta}2 \bar V_{N,\odd}}
		\Big)\bigg] . \label{Z_N 5}
	\end{align}
By choosing the observable operator $\hat O = 1$ in \eqref{Z_N 5}, we obtain the Araki--Lieb--Thirring inequality stated in \eqref{ALT N}:
\begin{equation*}
	\mathcal Z_N - \mathcal Z_{2N} = \frac12 \sqrt{\frac{m}{2\pi\beta}}\int_0^L \d x_0 \times
	\mathbb E_{\sigma_{2N}[x_0]}
	\bigg[\Big( e^{-\frac{\beta}2 \bar V_{N,\even}} - e^{-\frac{\beta}2 \bar V_{N,\odd}}
	\Big)^2\bigg] \Ge 0.
\end{equation*}
For confining potentials, the expression of $\mathcal W_N - \mathcal W_{2N}$ is the same as \eqref{Z_N 5} except that the integral domain is replaced by $\mathbb R$.

It remains to bound the right-hand side of \eqref{Z_N 5} pointwise in $x_0$. The even and odd terms cancel in the differences $\bar V_{N,\even} - \bar V_{N,\odd}$ and $\bar O_{N,\even} - \bar O_{N,\odd}$, so the contribution scales as $\mathcal O(1/N^2)$.

\section{Proofs of main results}
\label{section: proof}

This section bounds the successive difference $\mathcal W_N - \mathcal W_{2N}$ via the path integral representation \eqref{Z_N 5}, from which Theorems~\ref{theorem: periodic} and \ref{theorem: confining} follow. The analysis hinges on bounding the integrand of \eqref{Z_N 5}. Since the observable $O(x)$ is bounded and $V(x) \Ge 0$ under our assumptions, a direct application of elementary inequalities yields:
\begin{align*}
	& ~~~~ \bigg|\Big(
	\bar O_{N,\even} e^{-\frac{\beta}2 \bar V_{N,\even}} -
	\bar O_{N,\odd} e^{-\frac{\beta}2 \bar V_{N,\odd}}
	\Big)\Big( e^{-\frac{\beta}2 \bar V_{N,\even}} - e^{-\frac{\beta}2 \bar V_{N,\odd}}
	\Big)\bigg| \\
	& \Le 2K^2\Big(\big|
	\bar O_{N,\even} - \bar O_{N,\odd}
	\big|^2 + \big|
	e^{-\frac{\beta}2 \bar V_{N,\even}} - e^{-\frac{\beta}2 \bar V_{N,\odd}}
	\big|^2\Big) \\
	& \Le 2K^2\Big(\big|
	\bar O_{N,\even} - \bar O_{N,\odd}
	\big|^2 + \beta^2\big|
	\bar V_{N,\even} - \bar V_{N,\odd}
	\big|^2\Big) \\
	& = \frac{2K^2}{N^2}\Bigg[\bigg(\sum_{j=0}^{N-1} O(x_{2j}) - O(x_{2j+1})\bigg)^2 + \beta^2\bigg(\sum_{j=0}^{N-1} V(x_{2j}) - V(x_{2j+1})\bigg)^2\Bigg].
\end{align*}
For periodic potentials, substituting this result into \eqref{Z_N 5} gives the following bound:
	\begin{align}
		& |\mathcal W_N - \mathcal W_{2N}| \Le
		\frac{2K^2}{N^2} \int_0^L \d x_0 \,\times \notag \\
		& \mathbb E_{\sigma_{2N}[x_0]} 
		\Bigg[\bigg(\sum_{j=0}^{N-1} \big(O(x_{2j}) - O(x_{2j+1})\big)\bigg)^2 + \beta^2 \bigg(\sum_{j=0}^{N-1} \big(V(x_{2j}) - V(x_{2j+1})\big)\bigg)^2\Bigg].
		\label{Z_N 6}
	\end{align}
It thus remains to estimate the second moment
\begin{equation}
	\mathcal I(x_0) := \mathbb E_{\sigma_{2N}[x_0]}
	\Bigg[\bigg(
	\sum_{j=0}^{N-1} \big(F(x_{2j}) - F(x_{2j+1})\big)
	\bigg)^2\Bigg],
	\label{I x0}
\end{equation}
where $F$ stands for either the observable $O$ or the potential $V$, and $\sigma_{2N}[x_0]$ is the discrete Brownian bridge of \Cref{lemma: Brownian}.

\subsection{Second-moment bound on the Brownian bridge}

The bound on $\mathcal I(x_0)$ depends on the growth properties of the derivative $F'(x)$. The following lemma addresses the two scenarios relevant to our analysis: a uniformly bounded derivative (corresponding to periodic potentials) and a derivative with at most linear growth (corresponding to confining potentials).
\begin{lemma}
	\label{lemma: I bound}
	For the quantity $\mathcal I(x_0)$ defined in \eqref{I x0}:\\[-4pt]
	\begin{enumerate}[(i)]
		\setlength{\itemsep}{8pt}
		\item 
		If the function $F(x)$ satisfies
		\begin{equation*}
			\sup_{x\in\mathbb R} \big(|F'(x)| + |F''(x)|\big) \Le K,
		\end{equation*}
		then $\mathcal I(x_0) \Le \dfrac{25K^2\beta}{m}\bigg(\dfrac{\beta}{m}+1\bigg)$ for any $x_0\in\mathbb R$.
		\item
		If the function $F(x)$ satisfies
		\begin{equation*}
			\sup_{x\in\mathbb R} \bigg(
			\frac{|F'(x)|}{|x|+1} + |F''(x)|
			\bigg) \Le K,
		\end{equation*}
		then $\mathcal I(x_0) \Le \dfrac{80K^2\beta}{m}\bigg(\dfrac{\beta}{m}+1\bigg)(x_0^2+1)$ for any $x_0\in\mathbb R$.
	\end{enumerate}
\end{lemma}

\begin{proof}
	(i) For $0\Le j,l\Le N-1$, define the quantity 
	\begin{equation} 
		\mathcal I_{j,l} := \mathbb E_{\sigma_{2N}[x_0]} 
		\Big[ 
		\big(F(x_{2j}) - F(x_{2j+1})\big) 
		\big(F(x_{2l}) - F(x_{2l+1})\big) 
		\Big]. 
	\end{equation} 
	We estimate the diagonal and off-diagonal sums separately. For the diagonal sum, using $|F'(x)|\Le K$ and \Cref{lemma: Brownian},
	\begin{align*}
		\sum_{j=0}^{N-1} \mathcal I_{j,j} & =  
		\sum_{j=0}^{N-1} 
		\mathbb E_{\sigma_{2N}[x_0]} \Big[ 
		\big(F(x_{2j}) - F(x_{2j+1})\big)^2 
		\Big] \notag \\ 
		& \Le K^2 \sum_{j=0}^{N-1} \mathbb E_{\sigma_{2N}[x_0]}
		\big[(x_{2j+1}-x_{2j})^2\big] \Le K^2\sum_{j=0}^{N-1} \frac{\beta}{2mN} = \frac{K^2\beta}{2m}.
	\end{align*}
	For the off-diagonal sum, assume $j<l$. The indices involved in $\mathcal I_{j,l}$ are arranged as
	\begin{equation*} 
		2j < 2j+1 < 2l < 2l+1.
	\end{equation*} 
	By the Markov property of the discrete Brownian bridge, we simplify $\mathcal I_{j,l}$ by conditioning on $x_{2j+1}$ and $x_{2l}$. Using $|F''(x)|\Le K$ and Taylor expansion,
	\begin{subequations}
		\begin{align} 
			\big|F(x_{2j}) - F(x_{2j+1}) - 
			(x_{2j}-x_{2j+1}) F'(x_{2j+1})\big| & \Le  
			\frac{K}2(x_{2j}-x_{2j+1})^2,
			\label{VD a}
			\\ 
			\big|F(x_{2l+1}) - F(x_{2l}) - 
			(x_{2l+1}-x_{2l}) F'(x_{2l})\big| & \Le  
			\frac{K}2(x_{2l+1}-x_{2l})^2. 
			\label{VD b}
		\end{align} 
	\end{subequations}
	By \Cref{lemma: Brownian},
	\begin{align*} 
		\mathbb E_{\sigma_{2N}[x_0]}\big[(x_{2j}-x_{2j+1})\big|x_{2j+1}\big] & = -\frac{x_{2j+1}-x_0}{2j+1}, \\ 
		\mathbb E_{\sigma_{2N}[x_0]}\big[(x_{2j}-x_{2j+1})^2\big|x_{2j+1}\big] & = \frac{\beta j}{mN(2j+1)} + \frac{(x_{2j+1}-x_0)^2}{(2j+1)^2}, \\ 
		\mathbb E_{\sigma_{2N}[x_0]}\big[(x_{2l+1}-x_{2l})\big|x_{2l}\big] & =  
		-\frac{x_{2l}-x_0}{2N-2l}, \\ 
		\mathbb E_{\sigma_{2N}[x_0]}\big[(x_{2l+1}-x_{2l})^2\big|x_{2l}\big] & =  
		\frac{\beta(2N-2l-1)}{4mN(N-l)} + \frac{(x_{2l}-x_0)^2}{(2N-2l)^2}.
	\end{align*} 
	Hence taking the conditional expectation in \eqref{VD a} yields
	\begin{align}
		& ~~~~ \bigg|\mathbb E_{\sigma_{2N}[x_0]}\big[F(x_{2j}) - F(x_{2j+1})\big|x_{2j+1}\big] + 
		\frac{x_{2j+1}-x_0}{2j+1} F'(x_{2j+1})\bigg| \notag \\
		& \Le \frac{K}2 \mathbb E_{\sigma_{2N}[x_0]} \big[(x_{2j} - x_{2j+1})^2 | x_{2j+1}\big] \Le K\bigg(\frac{\beta}{2mN} + \frac{(x_{2j+1}-x_0)^2}{(2j+1)^2}\bigg).
		\label{FF bound}
	\end{align}
	Using $|F'(x)|\Le K$, we obtain
	\begin{equation}
		\Big|\mathbb E_{\sigma_{2N}[x_0]}\big[F(x_{2j}) - F(x_{2j+1})\big|x_{2j+1}\big]
		\Big| 
		\Le K\bigg(\frac{\beta}{2m N} + \frac{|x_{2j+1}-x_0|}{2j+1} + \frac{(x_{2j+1}-x_0)^2}{(2j+1)^2}\bigg).
		\label{EV 1}
	\end{equation}
	Similarly, for $x_{2l}$ and $x_{2l+1}$ in \eqref{VD b} we have the inequality
	\begin{equation}
		\Big|\mathbb E_{\sigma_{2N}[x_0]}\big[F(x_{2l}) - F(x_{2l+1})\big|x_{2l}\big]
		\Big| 
		\Le K\bigg(\frac{\beta}{2mN} + \frac{|x_{2l}-x_0|}{2N-2l} + \frac{(x_{2l}-x_0)^2}{(2N-2l)^2}\bigg).
		\label{EV 2}
	\end{equation}
	According to \Cref{lemma: Brownian}, the second moments of the Gaussian random variables $x_{2j+1}-x_0$ and $x_{2l}-x_0$ are bounded by 
	\begin{equation*} 
		\mathbb E_{\sigma_{2N}[x_0]}\big[(x_{2j+1}-x_0)^2\big] \Le \frac{\beta(2j+1)}{2mN},\qquad  
		\mathbb E_{\sigma_{2N}[x_0]}\big[(x_{2l}-x_0)^2\big] \Le \frac{\beta(2N-2l)}{2mN}.
	\end{equation*} 
	Multiplying \eqref{EV 1} and \eqref{EV 2} and taking the expectation over $x_{2j+1}$ and $x_{2l}$,
	\begin{align*}
		& ~~~~|\mathcal I_{j,l}| = \bigg|\mathbb E_{\sigma_{2N}[x_0]}
		\Big[
		\big(F(x_{2j}) - F(x_{2j+1})\big)
		\big(F(x_{2l}) - F(x_{2l+1})\big)
		\Big]\bigg| \\
		& \Le K^2\cdot \mathbb E\Bigg[
		\bigg(\frac{\beta}{2mN} + \frac{|x_{2j+1}-x_0|}{2j+1} + \frac{(x_{2j+1}-x_0)^2}{(2j+1)^2}\bigg)
		\bigg(\frac{\beta}{2mN} + \frac{|x_{2l}-x_0|}{2N-2l} + \frac{(x_{2l}-x_0)^2}{(2N-2l)^2}\bigg)\Bigg] \\
		& \Le
		3K^2\sqrt{ \mathbb E
			\bigg[\frac{\beta^2}{4m^2N^2} + \frac{(x_{2j+1}-x_0)^2}{(2j+1)^2} + \frac{(x_{2j+1}-x_0)^4}{(2j+1)^4}\bigg]
			\mathbb E
			\bigg[\frac{\beta^2}{4m^2N^2} + \frac{(x_{2l}-x_0)^2}{(2N-2l)^2} + \frac{(x_{2l}-x_0)^4}{(2N-2l)^4}\bigg]} \\
		& \Le 3K^2\sqrt{\bigg(\frac{\beta^2}{4m^2N^2} + \frac{\beta}{m(2j+1)N} + \frac{\beta^2}{m^2(2j+1)^2N^2}\bigg)
			\bigg(\frac{\beta^2}{4m^2N^2} + \frac{\beta}{m(2N-2l)N} + \frac{\beta^2}{m^2(2N-2l)^2N^2}\bigg)} \\
		& \Le \frac{6K^2\beta}m\bigg(\frac{\beta}m+1\bigg) \sqrt{\frac1{(2j+1)N} \cdot \frac1{(2N-2l)N}} =  \frac{6K^2\beta}m\bigg(\frac{\beta}m+1\bigg)\frac1{\sqrt{(2j+1)(2N-2l)}N}.
	\end{align*}
	As a consequence, the off-diagonal sum is bounded by
	$$
	\sum_{0\Le j<l\Le N-1}
	|\mathcal I_{j,l}| \Le \frac{6K^2\beta}{mN}\bigg(\frac{\beta}m+1\bigg) \sum_{j,l=0}^{N-1}
	\frac1{\sqrt{(2j+1)(2N-2l)}} \Le \frac{12K^2\beta}{m}\bigg(\frac{\beta}m+1\bigg),
	$$
	where we used $\sum_{j=0}^{N-1}\frac{1}{\sqrt{2j+1}} \Le \sqrt{2N}$ and $\sum_{l=0}^{N-1}\frac{1}{\sqrt{2N-2l}} \Le \sqrt{2N}$ to bound the double sum by $2N$. Combining the diagonal bound with the doubled off-diagonal contribution (counting both $j<l$ and $j>l$), $\mathcal I(x_0) \Le \big(\tfrac{1}{2} + 24\big)\tfrac{K^2\beta}{m}\big(\tfrac{\beta}{m}+1\big) \Le \tfrac{25K^2\beta}{m}\big(\tfrac{\beta}{m}+1\big)$. This completes the first part of \Cref{lemma: I bound}.\\[6pt]
	(ii) The argument parallels part~(i), with the linear-growth bound on $F'(x)$ introducing factors of $|x|+1$. We again split $\sum_{j,l}\mathcal I_{j,l}$ into diagonal and off-diagonal contributions.
	For the diagonal sum, applying the mean-value representation and Cauchy's inequality, we have
	\begin{align*} 
		\sum_{j=0}^{N-1} \mathcal I_{j,j} & =  
		\sum_{j=0}^{N-1} 
		\mathbb E_{\sigma_{2N}[x_0]} \Big[ 
		\big(F(x_{2j}) - F(x_{2j+1})\big)^2 
		\Big] \notag \\ 
		& = \sum_{j=0}^{N-1} \mathbb E_{\sigma_{2N}[x_0]} 
		\Bigg[\bigg((x_{2j+1}-x_{2j}) \int_0^1 F'\big(\theta x_{2j} + (1-\theta) x_{2j+1}\big)\d\theta\bigg)^2\Bigg] \\
		& \Le 3K^2 \sum_{j=0}^{N-1} \mathbb E_{\sigma_{2N}[x_0]} \Big[
		(x_{2j+1}-x_{2j})^2\big(x_{2j}^2+x_{2j+1}^2+1\big)\Big] \\
		& \Le 3K^2 \sum_{j=0}^{N-1} \sqrt{\mathbb E_{\sigma_{2N}[x_0]} \Big[(x_{2j+1}-x_{2j})^4\Big]
			\mathbb E_{\sigma_{2N}[x_0]} \Big[\big(x_{2j}^2 + x_{2j+1}^2 + 1\big)^2 \Big]
		}.
	\end{align*}
	Since $x_{2j}$ and $x_{2j+1}$ are Gaussian random variables, we obtain
	\begin{align*}
		\mathbb E_{\sigma_{2N}[x_0]} \Big[(x_{2j+1}-x_{2j})^4\Big] & \Le
		\frac{3\beta^2}{4m^2N^2}, \\
		\mathbb E_{\sigma_{2N}[x_0]} \Big[\big(x_{2j}^2 + x_{2j+1}^2 + 1\big)^2 \Big] & \Le
		9\bigg(\frac{\beta}{m}+1\bigg)^2(x_0^4+1).
	\end{align*}
	Hence the diagonal sum satisfies
	\begin{equation*}
		\sum_{j=0}^{N-1} \mathcal I_{j,j} \Le
		\frac{8K^2\beta}{m}\bigg(\frac{\beta}{m}+1\bigg)(x_0^2+1).
	\end{equation*}
	For the off-diagonal sum, assume $j<l$. Using the linear growth $|F'(x)|\Le K(|x|+1)$ together with \eqref{FF bound}, we obtain
	\begin{align}
		& \Big|\mathbb E_{\sigma_{2N}[x_0]}\big[F(x_{2j}) - F(x_{2j+1})\big|x_{2j+1}\big]
		\Big| \notag \\
		& \Le K\bigg(\frac{\beta}{2mN} + \frac{|x_{2j+1}-x_0|(|x_{2j+1}|+1)}{2j+1} + \frac{(x_{2j+1}-x_0)^2}{(2j+1)^2}\bigg). \label{EF 1}
	\end{align}
	Similarly, for $x_{2l}$ and $x_{2l+1}$ we have the inequality
	\begin{align}
		& \Big|\mathbb E_{\sigma_{2N}[x_0]}\big[F(x_{2l}) - F(x_{2l+1})\big|x_{2l}\big]
		\Big| \notag \\
		& \Le K\bigg(\frac{\beta}{2mN} + \frac{|x_{2l}-x_0|(|x_{2l}|+1)}{2N-2l} + \frac{(x_{2l}-x_0)^2}{(2N-2l)^2}\bigg). \label{EF 2}
	\end{align}
	We multiply \eqref{EF 1} and \eqref{EF 2} and take expectation. Applying  $(|x|+1)^2 \Le 3((x-x_0)^2+x_0^2+1)$ and the Gaussian moment estimates
	\begin{equation*}
		\mathbb E[(x_{2j+1}-x_0)^2]\Le \frac{\beta(2j+1)}{2mN}, \quad 
		\mathbb E[(x_{2j+1}-x_0)^4]\Le \frac{3\beta^2(2j+1)^2}{4m^2N^2},
	\end{equation*}
	we obtain
	\begin{align*}
		& ~~~~ |\mathcal I_{j,l}| = \bigg|\mathbb E_{\sigma_{2N}[x_0]}
		\Big[
		\big(F(x_{2j}) - F(x_{2j+1})\big)
		\big(F(x_{2l}) - F(x_{2l+1})\big)
		\Big]\bigg| \\
		& \Le K^2\cdot \mathbb E\Bigg[
		\bigg(\frac{\beta}{2mN} + \frac{|x_{2j+1}-x_0|(|x_{2j+1}|+1)}{2j+1} + \frac{(x_{2j+1}-x_0)^2}{(2j+1)^2}\bigg) \\
		& \hspace{1.6cm} \cdot \bigg(\frac{\beta}{2mN} + \frac{|x_{2l}-x_0|(|x_{2l}|+1)}{2N-2l} + \frac{(x_{2l}-x_0)^2}{(2N-2l)^2}\bigg)\Bigg] \\
		& \Le 3K^2\sqrt{\mathbb E\bigg[\frac{\beta^2}{4m^2N^2} + \frac{(x_{2j+1}-x_0)^2(|x_{2j+1}|+1)^2}{(2j+1)^2} + \frac{(x_{2j+1}-x_0)^4}{(2j+1)^4}\bigg]} \\
		& \hspace{1.6cm} \cdot \sqrt{\mathbb E\bigg[\frac{\beta^2}{4m^2N^2} + \frac{(x_{2l}-x_0)^2(|x_{2l}|+1)^2}{(2N-2l)^2} + \frac{(x_{2l}-x_0)^4}{(2N-2l)^4}\bigg]} \\
		& \Le 3K^2\sqrt{\frac{6\beta(x_0^2+1)}{m(2j+1)N}\bigg(\frac{\beta}m+1\bigg) \cdot \frac{6\beta(x_0^2+1)}{m(2N-2l)N}\bigg(\frac{\beta}m+1\bigg)} \\
		& = \frac{18K^2\beta(x_0^2+1)}{m\sqrt{(2j+1)(2N-2l)}\,N}\bigg(\frac{\beta}m+1\bigg).
	\end{align*}
	As a consequence, the off-diagonal sum is bounded by
	\begin{align*}
	\sum_{0\Le j<l\Le N-1}
	|\mathcal I_{j,l}| & \Le \frac{18K^2\beta(x_0^2+1)}{mN}\bigg(\frac{\beta}m+1\bigg)\sum_{j,l=0}^{N-1}
	\frac1{\sqrt{(2j+1)(2N-2l)}} \\
	& \Le \frac{36K^2\beta(x_0^2+1)}{m}\bigg(\frac{\beta}m+1\bigg),
	\end{align*}
	using the square-root sum bound $\sum_{j,l}1/\sqrt{(2j+1)(2N-2l)} \Le 2N$ as in part~(i). Combining the diagonal bound with the doubled off-diagonal contribution (counting both $j<l$ and $j>l$),
	$$
	\mathcal I(x_0) \Le \bigg(\frac{8K^2\beta}{m}\bigg(\frac{\beta}{m}+1\bigg) + 2\cdot\frac{36K^2\beta}{m}\bigg(\frac{\beta}m+1\bigg)\bigg)(x_0^2+1) \Le \frac{80K^2\beta}{m}\bigg(\frac{\beta}m+1\bigg)(x_0^2+1).
	$$
	This completes the second part of \Cref{lemma: I bound}.
\end{proof}

\subsection{Proof of \Cref{theorem: periodic}}

The error analysis for periodic potentials combines the integrand bound \eqref{Z_N 6} with the second-moment estimate from \Cref{lemma: I bound}.

Under Assumption~\ref{assumption: periodic}, both $O(x)$ and $V(x)$ satisfy $\sup_x(|F'(x)|+|F''(x)|) \Le K$, so part~(i) of \Cref{lemma: I bound} applies to either $F = O$ or $F = V$ and yields
\begin{equation*}
	\mathcal I_F(x_0) := \mathbb E_{\sigma_{2N}[x_0]}\bigg[\bigg(\sum_{j=0}^{N-1}\big(F(x_{2j}) - F(x_{2j+1})\big)\bigg)^2\bigg] \Le \frac{25K^2\beta}{m}\bigg(\frac{\beta}{m}+1\bigg).
\end{equation*}
Substituting into \eqref{Z_N 6} and integrating over $[0,L]$,
\begin{equation}
	|\mathcal W_N - \mathcal W_{2N}| \Le \frac{2K^2}{N^2}\int_0^L\big(\mathcal I_O(x_0) + \beta^2\mathcal I_V(x_0)\big)\d x_0 \Le \frac{50K^4 L\beta(1+\beta^2)}{mN^2}\bigg(\frac{\beta}{m}+1\bigg).
	\label{periodic: 1}
\end{equation}
Writing the total error as the telescoping series $\mathcal W_N - \mathcal W = \sum_{j=0}^\infty(\mathcal W_{2^jN} - \mathcal W_{2^{j+1}N})$ and applying \eqref{periodic: 1} to each term,
\begin{align*}
	|\mathcal W_N - \mathcal W| & \Le \sum_{j=0}^\infty
	|\mathcal W_{2^{j}N} - \mathcal W_{2^{j+1}N}| \Le
	\sum_{j=0}^\infty \frac{50K^4 L\beta(1+\beta^2)(\beta/m+1)}{m(2^j N)^2} \\
	& = \frac{200K^4 L\beta(1+\beta^2)}{3mN^2}\bigg(\frac{\beta}{m}+1\bigg),
\end{align*}
using $\sum_{j=0}^\infty 4^{-j} = 4/3$.
This establishes the $\mathcal O(1/N^2)$ convergence rate and proves \Cref{theorem: periodic}.

\subsection{Proof of \Cref{theorem: confining}}
\label{section: truncation}

The confining case poses two difficulties: the path integral extends over $\mathbb R$, and the confining structure of $V$ must be used to keep the integral well-defined.

We employ a \emph{domain truncation}: approximate the trace by integrating over $[-R, R]$ and bound the error introduced by this restriction. Analogous to the path integral formulation for the periodic case in \eqref{trace N PI}, we define the truncated trace quantity $\mathcal W_N(R)$ by restricting the base-point integral to this finite domain:
\begin{align}
	\mathcal W_N(R) & := \frac1N\sum_{j=0}^{N-1}\int_{-R}^R
	\mel{x_0}{\big(e^{-\frac{\beta}N \hat H_0}e^{-\frac{\beta}N \hat V}\big)^{N-j} \hat O\big(e^{-\frac{\beta}N \hat H_0}e^{-\frac{\beta}N \hat V}\big)^j}{x_0} \d x_0
	\notag \\
	& = \sqrt{\frac{m}{2\pi\beta}}\int_{-R}^R \d x_0
	\times \mathbb E_{\sigma_N[x_0]}
	\bigg[
	\frac1N\sum_{j=0}^{N-1} O(x_j) \times
	\exp\bigg(-\frac{\beta}{N}
	\sum_{j=0}^{N-1} V(x_j)\bigg)
	\bigg].
	\label{W_N 1}
\end{align}
The definition of $\mathcal W_N(R)$ in \eqref{W_N 1} involves an explicit average of the observable over all $N$ time slices. This averaging is necessary because integrating over a finite domain $[-R, R]$ breaks the cyclic symmetry of the trace.
In the limit as $R\to\infty$, the integration domain expands to $\mathbb R$, which restores the cyclic symmetry, and the expression for $\mathcal W_N(R)$ converges to the untruncated quantity $\mathcal W_N$:
\begin{align*}
	\lim_{R\to\infty} \mathcal W_N(R) & = \frac1N\sum_{j=0}^{N-1} \int_{\mathbb R}
	\mel{x_0}{\big(e^{-\frac{\beta}N \hat H_0}e^{-\frac{\beta}N \hat V}\big)^{N-j} \hat O\big(e^{-\frac{\beta}N \hat H_0}e^{-\frac{\beta}N \hat V}\big)^j}{x_0} \d x_0 \\
	& = \int_{\mathbb R}
	\mel{x_0}{\big(e^{-\frac{\beta}N \hat H_0}e^{-\frac{\beta}N \hat V}\big)^{N} \hat O}{x_0} \d x_0 = \tr\Big[\big(e^{-\frac{\beta}N \hat H_0}e^{-\frac{\beta}N \hat V}\big)^{N}\hat O\Big] = \mathcal W_N.
\end{align*}

To facilitate the error analysis, we introduce two distinct path integral representations for the truncated trace based on a $2N$-step path. First, in the even-steps representation, both the observable and the potential are evaluated on the even-indexed path variables. This is analogous to the untruncated expression in \eqref{Z_N 1}:
\begin{align}
	\mathcal W_N(R)  = \sqrt{\frac{m}{2\pi\beta}} & \int_{-R}^R \d x_0 
	\times \mathbb E_{\sigma_{2N}[x_0]}
	\bigg[
	\frac1N\sum_{j=0}^{N-1} O(x_{2j})
	\times
	\exp\bigg(-\frac{\beta}{N}
	\sum_{j=0}^{N-1} V(x_{2j})\bigg)
	\bigg] \notag \\
	& = \sqrt{\frac{m}{2\pi\beta}}\int_{-R}^R 
	\d x_0 
	\times \mathbb E_{\sigma_{2N}[x_0]}
	\Big[\bar O_{N,\even}\times e^{-\beta \bar V_{N,\even}}\Big].
	\label{Z_N R 1}
\end{align}
Second, we define a corresponding odd-steps representation, where both the observable and potential are evaluated on the odd-indexed variables:
\begin{align}
	\tilde{\mathcal W}_N(R) = \sqrt{\frac{m}{2\pi\beta}} & \int_{-R}^R \d x_0 
	\times \mathbb E_{\sigma_{2N}[x_0]}
	\bigg[
	\frac1N\sum_{j=0}^{N-1} O(x_{2j+1})
	\times
	\exp\bigg(-\frac{\beta}{N}
	\sum_{j=0}^{N-1} V(x_{2j+1})\bigg)
	\bigg] \notag \\
	& =  \sqrt{\frac{m}{2\pi\beta}}\int_{-R}^R 
	\d x_0 
	\times \mathbb E_{\sigma_{2N}[x_0]}
	\Big[\bar O_{N,\odd}\times e^{-\beta \bar V_{N,\odd}}\Big].
	\label{Z_N R 2}
\end{align}
While $\mathcal W_N(R)$ and $\tilde{\mathcal W}_N(R)$ agree in the limit $R\to\infty$, the finite domain breaks the cyclic symmetry, and for any finite $R>0$ they need not coincide.

The following lemma establishes that the error introduced by this domain truncation decays exponentially with $R$.
\begin{lemma}
	\label{lemma: Z error}
	Under Assumption~\ref{assumption: confining}, the truncation errors are bounded by
	\begin{equation*}
		|\mathcal W_N(R) - \mathcal W_N|
		\Le \frac{C_R}{R} e^{-cR^2},\qquad
		|\tilde {\mathcal W}_N(R) - \mathcal W_N|
		\Le \frac{C_R}{R} e^{-cR^2},
	\end{equation*}
	where $c = \dfrac{a\beta}{2(1+a\beta^2/(4m))}$ and $C_R = \dfrac{K}{c}\sqrt{\dfrac{m}{2\pi\beta}}$.
\end{lemma}
\begin{proof}
	Throughout this proof, $C_{\beta,m}$ denotes a positive constant depending on $\beta$, $m$, and the assumption constants, whose value may change from line to line. The truncation error is the integral over the exterior domain
	$$
	[-R,R]^c = (-\infty,-R)\cup(R,\infty).
	$$
	Using the boundedness of $O(x)$ and the quadratic lower bound on $V(x)$ from Assumption~\ref{assumption: confining}, we can bound the error for the even-steps representation:
	\begin{align}
		|\mathcal W_N(R) - \mathcal W_N| & \Le C_{\beta,m}
		\int_{[-R,R]^c} \d x_0 \times 
		\mathbb E_{\sigma_{2N}[x_0]}
		\bigg[
		\exp\bigg(-\frac{\beta}{N}
		\sum_{j=0}^{N-1} V(x_{2j}) \bigg)
		\bigg] \notag \\
		& \Le C_{\beta,m}
		\int_{[-R,R]^c} \d x_0 \times 
		\mathbb E_{\sigma_{2N}[x_0]}
		\bigg[
		\exp\bigg(-\frac{a\beta}{2N}
		\sum_{j=0}^{N-1} x_{2j}^2\bigg)
		\bigg] \notag \\
		& \Le C_{\beta,m} \int_R^\infty \d x_0 \times 
		\bigg(
		\frac1N \sum_{j=0}^{N-1}
		\mathbb E_{\sigma_{2N}[x_0]}
		\Big[
		e^{-\frac{a\beta}{2} x_{2j}^2}
		\Big]\bigg),
		\label{WRW difference N}
	\end{align}
	where we applied Jensen's inequality to the convex exponential function in the final step. The path variables $\{x_j\}_{j=0}^{2N-1}$ are distributed as a discrete Brownian bridge, so the marginal distribution of any $x_{2j}$ conditioned on $x_0$ is Gaussian $\mathcal N(x_0,\Sigma_{2j})$. Its variance, $\Sigma_{2j} = \frac{\beta}{m}\frac{j(N-j)}{N^2}$, is uniformly bounded by $\frac{\beta}{4m}$. A direct computation of the expectation then yields a uniform Gaussian bound in $x_0$:
	\begin{equation*}
		\mathbb E_{\sigma_{2N}[x_0]}
		\big[
		e^{-\frac{a\beta}{2} x_{2j}^2}
		\big] \Le \exp\bigg(-\frac{a\beta x_0^2}{2(1+a\beta^2/(4m))}\bigg).
	\end{equation*}
	Substituting this back and using the bound for the Gaussian tail gives
	\begin{equation}
		|\mathcal W_N(R) - \mathcal W_N| \Le 
		C_{\beta,m}\int_R^\infty \exp\bigg(-\frac{a\beta x_0^2}{2(1+a\beta^2/(4m))}\bigg)
		\d x_0 \Le 
		\frac{C_R}{R} e^{-cR^2}.
	\end{equation}
	An identical analysis holds for the odd-steps representation, $|\tilde {\mathcal W}_N(R) - \mathcal W_N|$, as the odd-indexed path variables $x_{2j+1}$ have the same distributional properties.
\end{proof}

The analysis for confining potentials combines the domain truncation method with the error bounds derived previously. Using the path integral expressions from \eqref{Z_N R 1} and \eqref{Z_N R 2}, we derive an equality analogous to \eqref{Z_N 5}:
	\begin{align*}
		& \frac12\big(\mathcal W_N(R) + \tilde{\mathcal W}_N(R)\big) - \mathcal W_{2N}(R) = \frac12 \sqrt{\frac{m}{2\pi\beta}}\int_{-R}^R \d x_0
		\,\times \\
		\mathbb E_{\sigma_{2N}[x_0]} &
		\bigg[\Big(
		\bar O_{N,\even} e^{-\frac{\beta}2 \bar V_{N,\even}} -
		\bar O_{N,\odd} e^{-\frac{\beta}2 \bar V_{N,\odd}}
		\Big)\Big( e^{-\frac{\beta}2 \bar V_{N,\even}} - e^{-\frac{\beta}2 \bar V_{N,\odd}}
		\Big)\bigg].
	\end{align*}
Following the procedure that produced \eqref{Z_N 6} in the periodic case, we obtain
	\begin{align}
		& \bigg|\frac12\big(\mathcal W_N(R) + \tilde{\mathcal W}_N(R)\big) - \mathcal W_{2N}(R)\bigg| \Le
		\frac{2K^2}{N^2} \int_{-R}^R \d x_0 \,\times
		\notag
		\\
		\mathbb E_{\sigma_{2N}[x_0]} &
		\Bigg[\bigg(\sum_{j=0}^{N-1} \big(O(x_{2j}) - O(x_{2j+1})\big)\bigg)^2 + \beta^2\bigg(\sum_{j=0}^{N-1} \big(V(x_{2j}) - V(x_{2j+1})\big)\bigg)^2\Bigg].
		\label{Z_N 7}
	\end{align}
Under Assumption~\ref{assumption: confining}, both $O$ and $V$ satisfy $\sup_x\big(|F'(x)|/(|x|+1)+|F''(x)|\big)\Le K$, so part~(ii) of \Cref{lemma: I bound} yields
\begin{equation*}
	\mathcal I_F(x_0) \Le \frac{80K^2\beta(x_0^2+1)}{m}\bigg(\frac{\beta}{m}+1\bigg) \qquad \text{for } F\in\{O,V\}.
\end{equation*}
Substituting into \eqref{Z_N 7} and using $\int_{-R}^R(x_0^2+1)\d x_0 \Le 2(R^3+R)$,
\begin{equation*}
	\bigg|\frac12\big(\mathcal W_N(R) + \tilde{\mathcal W}_N(R)\big) - \mathcal W_{2N}(R)\bigg| \Le \frac{320 K^4 \beta(1+\beta^2)(\beta/m+1)(R^3+R)}{mN^2}.
\end{equation*}
Combining with the truncation errors from \Cref{lemma: Z error} via the triangle inequality,
\begin{equation}
	|\mathcal W_N - \mathcal W_{2N}| \Le \frac{320 K^4 \beta(1+\beta^2)(\beta/m+1)(R^3+R)}{mN^2} + \frac{3C_R}{R}\,e^{-cR^2}.
	\label{Z_N bound R final}
\end{equation}
The choice $R^2 = \frac{2}{c}(\log N+1)$ (for $N$ large enough that $R \Ge 1$) gives $e^{-cR^2} = e^{-2}/N^2$ and $R^3+R \Le 2R^3 = \frac{4\sqrt 2(\log N+1)^{\frac32}}{c^{\frac32}}$. Hence
\begin{equation*}
	|\mathcal W_N - \mathcal W_{2N}| \Le \frac{C(\log N+1)^{\frac32}}{N^2}, \qquad C := \frac{1280\sqrt 2\, K^4\beta(1+\beta^2)(\beta/m+1)}{m\, c^{\frac32}} + 3C_R.
\end{equation*}
Summing the geometric series via $\sum_{j=0}^\infty (\log(2^jN)+1)^{\frac32}/4^j \Le 3(\log N+1)^{\frac32}$,
\begin{equation}
	|\mathcal W_N - \mathcal W| \Le \sum_{j=0}^\infty |\mathcal W_{2^jN} - \mathcal W_{2^{j+1}N}| \Le \frac{3C(\log N+1)^{\frac32}}{N^2}.
\end{equation}
This establishes the nearly optimal $\mathcal O\big((\log N+1)^{\frac32}/N^2\big)$ convergence rate and proves \Cref{theorem: confining}. The constant $3C$ depends on $a, \beta, m, K$ via $c$ and $C_R$ from \Cref{lemma: Z error}, and is independent of $N$.

\section{Numerical verification}
\label{section: numerical}

To corroborate the convergence rate of \Cref{theorem: confining}, we report a numerical test for the confining case. We take the smooth, confining potential
\begin{equation*}
	V(x) = \tfrac{1}{2}x^2 + 4\cos(x - 0.2),
\end{equation*}
the smooth, bounded observable
\begin{equation*}
	O(x) = -2\, e^{-x^2},
\end{equation*}
mass $m=1$, and the four inverse temperatures $\beta \in \{1,2,3,4\}$. The pair $(V, O)$ satisfies \Cref{assumption: confining}: $V$ has at least quadratic growth (for some small $a>0$), $V''$ and $|V'(x)|/(|x|+1)$ are bounded, and $O$ together with its first two derivatives is bounded.

The reference $\avg{\hat O}=\mathcal W/\mathcal Z$ is computed by full spectral diagonalization of $\hat H$ on a Fourier discrete-variable-representation grid of $N_x=256$ points on $[-12,12]$; the spatial truncation error is of spectral order and far below the Lie--Trotter signal at every $N$ considered. The Lie--Trotter approximant $\avg{\hat O}_N = \mathcal W_N/\mathcal Z_N$ is computed by representing the symmetric Lie--Trotter step $e^{-\frac{\beta}{2N}\hat V}e^{-\frac{\beta}{N}\hat H_0}e^{-\frac{\beta}{2N}\hat V}$ as a matrix on the same grid and raising it to the $N$-th power by repeated squaring. The number of Lie--Trotter steps ranges over $N = 2, 4, 8, \dots, 2^{12}$.

\begin{figure}[htbp]
	\centering
	\includegraphics[width=0.65\textwidth]{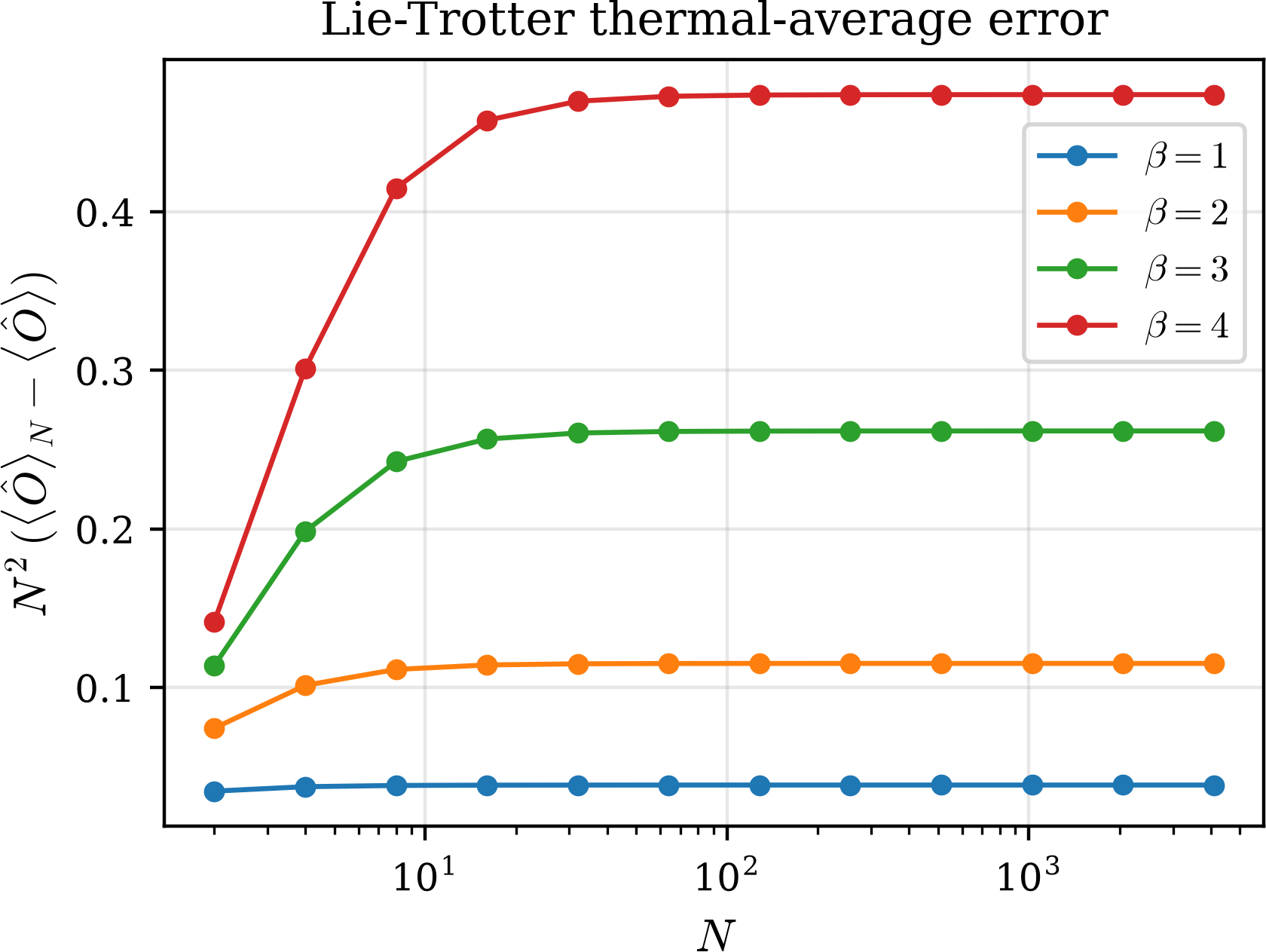}
	\caption{Rescaled Lie--Trotter error of the thermal average, $N^2(\avg{\hat O}_N - \avg{\hat O})$, against $N$ on a logarithmic scale, for the confining potential $V(x)=x^2/2 + 4\cos(x-0.2)$ and observable $O(x) = -2e^{-x^2}$ at inverse temperatures $\beta\in\{1,2,3,4\}$.}
	\label{fig: numerical}
\end{figure}

\Cref{fig: numerical} plots the rescaled error $N^2 (\avg{\hat O}_N - \avg{\hat O})$ as a function of $N$. Two features stand out. First, for each $\beta$ the rescaled error converges to a finite constant as $N$ increases, with the plateau already reached around $N\approx 64$ and remaining flat out to $N = 2^{12}$. No logarithmic enhancement is visible, suggesting that the $(\log N+1)^{\frac32}$ factor of \Cref{theorem: confining} is likely an artifact of the proof technique. Second, the asymptotic constant grows monotonically with $\beta$, from $\approx 4\times 10^{-2}$ at $\beta=1$ to $\approx 5\times 10^{-1}$ at $\beta=4$, consistent with the intuition that sampling at low temperature (large $\beta$) is more difficult.

\section{Conclusion}

This paper provides a non-asymptotic error analysis for the Lie--Trotter approximation in quantum statistical mechanics, addressing the problem of bounding the error for systems with unbounded Hamiltonians. Our analysis, conducted within a path integral framework, establishes an optimal $\mathcal O(1/N^2)$ convergence rate for both the partition function and related thermal averages in systems with smooth, periodic potentials. A key element of our proof is a path integral interpretation of the Araki--Lieb--Thirring inequality, which transforms the error analysis into the problem of bounding the difference between successive approximations.

We then extend this framework to the more challenging case of a confining potential on $\mathbb R$. Two difficulties arise: the unbounded integration domain and the linear growth of the potential's derivative. By employing a domain truncation strategy that balances the algebraic approximation error with an exponential truncation error, we derive a nearly optimal convergence rate of $\mathcal O((\log N+1)^{\frac32}/N^2)$.

These results provide quantitative error bounds for the Lie--Trotter approximation underlying PIMC and PIMD. Three possible directions for future work include removing the $(\log N+1)^{\frac32}$ artifact in higher dimensions, obtaining analogous error bounds for the real-time Schr\"odinger operator $e^{-it\hat H}$, and extending the analysis to high-order Suzuki product formulas \cite{suzuki1991general,hatano2005finding}.

\appendix

\section{Proof of \Cref{lemma: Brownian}}
\label{appendix: lemma proof}

The proof proceeds in three steps: first, we establish the joint distribution of $(x_1,\dots,x_{N-1})$; second, we derive the unconditional moments $\mathbb E\big[x_j^2\big]$ and $\mathbb E\big[(x_{j+1}-x_j)^2\big]$; third, we compute the conditional means $\mathbb E\big[x_{j\pm 1}\big|x_j\big]$ and conditional second moments $\mathbb E\big[(x_{j\pm 1}-x_j)^2\big|x_j\big]$.

\subsection{Joint distribution}
Specializing the density \eqref{sigma_N definition} to $x_0=0$ and enforcing the closure $x_N \equiv 0$ gives
\begin{equation*}
	\sigma_N[0](x_1, \dots, x_{N-1}) \propto \exp\bigg(- \frac{mN}{2\beta} \sum_{j=0}^{N-1} (x_{j+1}-x_j)^2\bigg),
\end{equation*}
with the boundary values fixed at $x_0 = x_N = 0$. Writing $\mathbf{x} = (x_1, \dots, x_{N-1})^\T \in \mathbb R^{N-1}$, the exponent is a quadratic form in $\mathbf{x}$, so $\sigma_N[0]$ is a centered multivariate Gaussian. Explicitly, the quadratic form factors as
$$
\frac{mN}{2\beta} \sum_{j=0}^{N-1} (x_{j+1}-x_j)^2 =
\frac{mN}{2\beta}\mathbf{x}^\T \mathbf{A} \mathbf{x},
$$
with $\mathbf{A} \in \mathbb R^{(N-1)\times(N-1)}$ the tridiagonal matrix
\begin{equation*}
	\mathbf{A} =
	\begin{pmatrix}
		2 & -1 & & \\
		-1 & 2 & -1 & \\
		& \ddots & \ddots & \ddots \\
		& & -1 & 2 & -1 \\
		& & & -1 & 2
	\end{pmatrix}.
\end{equation*}
The covariance matrix of $\sigma_N[0]$ is therefore
$$
\mathbf{\Sigma} = \bigg(\frac{mN}{\beta}\mathbf{A}\bigg)^{-1} = \frac{\beta}{mN}\mathbf{A}^{-1},
$$
and the closed-form inverse of $\mathbf{A}$,
\begin{equation*}
	(\mathbf{A}^{-1})_{jl} = \frac{\min(j,l)\big(N-\max(j,l)\big)}{N},
\end{equation*}
yields the entries
\begin{equation}
	\text{Cov}(x_j, x_l) = \mathbf{\Sigma}_{jl} = \frac{\beta}{m} \frac{\min(j,l)\big(N-\max(j,l)\big)}{N^2}.
	\label{eq:proof_cov}
\end{equation}

On the other hand, the process
$$
y_j = \sqrt{\frac{\beta}{m}}\bigg(B_{j/N} - \frac{j}{N} B_1\bigg),\qquad j=1,\dots,N-1,
$$
is a centered Gaussian process whose covariance, for $j \Le l$, is
\begin{align*}
	\text{Cov}(y_j, y_l) & = \frac{\beta}{m}\, \text{Cov}\bigg(B_{j/N} - \frac{j}{N} B_1,\ B_{l/N} - \frac{l}{N} B_1\bigg) \\
	& = \frac{\beta}{m} \bigg( \text{Cov}(B_{j/N}, B_{l/N}) - \frac{l}{N}\text{Cov}(B_{j/N}, B_1) - \frac{j}{N}\text{Cov}(B_1, B_{l/N}) + \frac{jl}{N^2}\text{Var}(B_1) \bigg) \\
	& = \frac{\beta}{m} \bigg( \frac{j}{N} - \frac{lj}{N^2} - \frac{jl}{N^2} + \frac{jl}{N^2} \bigg) = \frac{\beta}{m} \frac{j(N-l)}{N^2},
\end{align*}
which coincides with \eqref{eq:proof_cov} for $j\Le l$. Since $\{x_j\}_{j=1}^{N-1}$ and $\{y_j\}_{j=1}^{N-1}$ are centered Gaussians with the same covariance, they share the same joint law, proving the first part of the lemma.

\subsection{Unconditional moments}
We now compute $\mathbb E\big[x_j^2\big]$ and $\mathbb E\big[(x_{j+1}-x_j)^2\big]$; both follow directly from the covariance \eqref{eq:proof_cov}. The variance of $x_j$ is
\begin{equation*}
	\mathbb E\big[x_j^2\big] = \text{Cov}(x_j, x_j) = \frac{\beta}{m} \frac{j(N-j)}{N^2},
\end{equation*}
and the second moment of the increment $x_{j+1}-x_j$ is
\begin{align*}
	\mathbb E\big[(x_{j+1}-x_j)^2\big] &= \text{Var}(x_{j+1}) + \text{Var}(x_j) - 2\,\text{Cov}(x_{j+1}, x_j) \\
	&= \frac{\beta}{mN^2} \Big( (j+1)(N-j-1) + j(N-j) - 2j(N-j-1) \Big) \\
	&= \frac{\beta}{m} \frac{N-1}{N^2},
\end{align*}
which establishes the second part of the lemma.

\subsection{Conditional expectations}
We now compute the conditional means $\mathbb E\big[x_{j\pm 1}\big|x_j\big]$ and the conditional second moments $\mathbb E\big[(x_{j\pm 1}-x_j)^2\big|x_j\big]$. Let $(X, Y)$ be a centered jointly Gaussian pair. Then the conditional mean and variance of $X$ given $Y$ are
$$
\mathbb E[X|Y=y] = \frac{\text{Cov}(X,Y)}{\text{Var}(Y)}\, y, \qquad
\text{Var}(X|Y) = \text{Var}(X) - \frac{\text{Cov}(X,Y)^2}{\text{Var}(Y)}.
$$
Specializing these formulas to the pairs $(X,Y) = (x_{j-1}, x_j)$ and $(X,Y) = (x_{j+1}, x_j)$ and substituting \eqref{eq:proof_cov} give
\begin{align*}
	\mathbb E[x_{j-1}|x_j] &= \frac{j-1}{j}\, x_j,
	& \mathbb E[x_{j+1}|x_j] &= \frac{N-j-1}{N-j}\, x_j, \\
	\text{Var}(x_{j-1}|x_j) &= \frac{\beta}{m}\frac{j-1}{jN},
	& \text{Var}(x_{j+1}|x_j) &= \frac{\beta}{m}\frac{N-j-1}{N(N-j)},
\end{align*}
where the conditional variances follow from the algebraic simplification
$$
\frac{\beta}{m}\frac{(j-1)(N-j+1)}{N^2} - \frac{\big(\frac{\beta}{m}\frac{(j-1)(N-j)}{N^2}\big)^2}{\frac{\beta}{m}\frac{j(N-j)}{N^2}} = \frac{\beta}{m}\frac{j-1}{jN},
$$
and its $x_{j+1}$ analogue. Combining these with the identity
$$
\mathbb E\big[(X-Y)^2\big|Y=y\big] = \text{Var}(X|Y) + \big(\mathbb E[X|Y]-y\big)^2
$$
yields the conditional second moments
\begin{align*}
	\mathbb E\big[(x_{j-1}-x_j)^2\big|x_j\big] &= \frac{\beta}{m}\frac{j-1}{jN} + \frac{x_j^2}{j^2}, \\
	\mathbb E\big[(x_{j+1}-x_j)^2\big|x_j\big] &= \frac{\beta}{m}\frac{N-j-1}{N(N-j)} + \frac{x_j^2}{(N-j)^2},
\end{align*}
which completes the proof of the lemma.

\section{Proof of \Cref{lemma: consistency}}
\label{appendix: consistency}

Our goal is to bound $|\mathcal W_N - \mathcal W|$. Recall that for the periodic potential $V(x)$, the trace quantities $\mathcal W_N$ and $\mathcal W$ are given by \eqref{trace N PI} and \eqref{trace PI}:
\begin{align*}
	\mathcal W_N & = \sqrt{\frac{m}{2\pi\beta}}\int_0^L \d x_0
	\times \mathbb E_{\sigma[x_0]}
	\bigg[ \frac1N \sum_{j=0}^{N-1} O\Big(x\Big(\frac{j}{N}\Big)\Big)\times
	\exp\bigg(-\frac{\beta}{N}
		\sum_{j=0}^{N-1} V\Big(x\Big(\frac{j}{N}\Big)\Big)\bigg)
	\bigg], \\
	\mathcal W & = \sqrt{\frac{m}{2\pi\beta}}\int_0^L \d x_0
	\times \mathbb E_{\sigma[x_0]}
	\bigg[\int_0^1 O(x(\tau))\d\tau \times
	\exp\bigg(-\beta\int_0^1 V(x(\tau))\d\tau\bigg)
	\bigg],
\end{align*}
where $\sigma[x_0]$ is the continuous Brownian bridge of \eqref{continuous Brownian}. To keep the displays compact, we abbreviate the discrete $N$-bead mean and the continuous path average of a function $F$ along the path as
\begin{equation*}
	\bar F_N := \frac{1}{N}\sum_{j=0}^{N-1} F\Big(x\Big(\frac{j}{N}\Big)\Big), \qquad
	\bar F_\infty := \int_0^1 F(x(\tau))\d\tau,
\end{equation*}
so the integrands of $\mathcal W_N$ and $\mathcal W$ above read $\bar O_N\,e^{-\beta \bar V_N}$ and $\bar O_\infty\,e^{-\beta \bar V_\infty}$.

Under either Assumption~\ref{assumption: periodic} or Assumption~\ref{assumption: confining}, $O$ is bounded and $V$ is bounded below, so the Boltzmann factors $e^{-\beta \bar V_N}$ and $e^{-\beta \bar V_\infty}$ are uniformly bounded. Applying the triangle inequality to the difference of integrands yields
\begin{equation*}
	\big|\bar O_N\,e^{-\beta \bar V_N} - \bar O_\infty\,e^{-\beta \bar V_\infty}\big| \Le C_{\beta,m}\big(|\bar O_N - \bar O_\infty| + |\bar V_N - \bar V_\infty|\big),
\end{equation*}
where $C_{\beta,m}>0$ denotes a generic constant depending on $\beta$ and $m$, whose value may change from line to line. Integrating over $x_0$ then gives
\begin{equation}
	|\mathcal W_N - \mathcal W| \Le C_{\beta,m}\int_0^L \d x_0
	\times \mathbb E_{\sigma[x_0]}\big[|\bar O_N - \bar O_\infty| + |\bar V_N - \bar V_\infty|\big].
	\label{WW bound}
\end{equation}
Bounding $|\mathcal W_N - \mathcal W|$ thus reduces to controlling $\mathbb E_{\sigma[x_0]}[|\bar O_N - \bar O_\infty|]$ and $\mathbb E_{\sigma[x_0]}[|\bar V_N - \bar V_\infty|]$. These two quantities share the same structure---only the function being averaged ($O$ or $V$) differs---so we treat them uniformly by introducing a generic function $F$ and setting
\begin{equation}
	\mathcal J(x_0) = \mathbb E_{\sigma[x_0]}\big[|\bar F_N - \bar F_\infty|\big] = \mathbb E_{\sigma[x_0]}\Bigg[\bigg|\frac{1}{N}\sum_{j=0}^{N-1} F\Big(x\Big(\frac{j}{N}\Big)\Big) - \int_0^1 F(x(\tau))\d\tau\bigg|\Bigg],
	\label{J x0}
\end{equation}
where $F$ stands for either $V$ or $O$; $\mathcal J(x_0)$ is the first-moment analogue of $\mathcal I(x_0)$ from \eqref{I x0}. By \eqref{WW bound}, bounding $\mathcal J(x_0)$ uniformly for $F\in\{O,V\}$ is sufficient to estimate $|\mathcal W_N - \mathcal W|$. This is the content of \Cref{lemma: J bound}.

\subsection{First-moment bound on the Brownian bridge}
We now record the first-moment counterpart of \Cref{lemma: I bound}. Since only convergence is needed, the estimate uses only the first derivative of $F$.
\begin{lemma}
	\label{lemma: J bound}
	Let $\mathcal J(x_0)$ be defined as in \eqref{J x0}.
	\begin{enumerate}[(i)]
		\setlength{\itemsep}{8pt}
		\item If $\sup_{x\in\mathbb R} |F'(x)| <+\infty$, then
		$$
		\mathcal J(x_0)\Le \frac{C_{\beta,m}}{\sqrt{N}} \qquad \text{for every } x_0\in\mathbb R.
		$$
		\item If $\sup_{x\in\mathbb R} |F'(x)|/(|x|+1) <+\infty$, then
		$$
		\mathcal J(x_0) \Le \frac{C_{\beta,m}(|x_0|+1)}{\sqrt{N}} \qquad \text{for every } x_0\in\mathbb R.
		$$
	\end{enumerate}
	The constant $C_{\beta,m}$ is independent of $N$ and $x_0$, but may still depend on $\beta$, $m$, and the bounds on $F$ and $F'$.
\end{lemma}
\begin{proof}
	Writing each summand as an integral over $\tau\in[0,1]$ and applying the triangle inequality and Fubini's theorem,
	\begin{align}
		\mathcal J(x_0) & = \frac1N\mathbb E_{\sigma[x_0]} \Bigg[\bigg|
		\sum_{j=0}^{N-1} \bigg(F\Big(x\Big(\frac{j}{N}\Big)\Big) -
		\int_0^1 F\Big(x\Big(\frac{j+\tau}{N}\Big)\Big)\d\tau\bigg)
		\bigg|
		\Bigg] \notag \\
		& \Le \frac1N \sum_{j=0}^{N-1} \int_0^1 \mathbb E_{\sigma[x_0]}\Bigg[\bigg|
		F\Big(x\Big(\frac{j}{N}\Big)\Big) - F\Big(x\Big(\frac{j+\tau}{N}\Big)\Big)
		\bigg|\Bigg] \d\tau.
		\label{J x0 bound}
	\end{align}
	(i) If $|F'(x)|$ is uniformly bounded, the mean-value theorem gives
	$$
	\bigg|
	F\Big(x\Big(\frac{j}{N}\Big)\Big) - F\Big(x\Big(\frac{j+\tau}{N}\Big)\Big)
	\bigg| \Le C_{\beta,m}\Big|x\Big(\frac{j}{N}\Big) - x\Big(\frac{j+\tau}{N}\Big)\Big|.
	$$
	Taking expectations and using the standard Brownian-bridge increment bound,
	\begin{equation*}
		\mathbb E_{\sigma[x_0]}\Bigg[\bigg|
		F\Big(x\Big(\frac{j}{N}\Big)\Big) - F\Big(x\Big(\frac{j+\tau}{N}\Big)\Big)
		\bigg|\Bigg] \Le \frac{C_{\beta,m}}{\sqrt{N}},
	\end{equation*}
	and \eqref{J x0 bound} yields $\mathcal J(x_0) \Le C_{\beta,m}/\sqrt{N}$, establishing part (i).\\[6pt]
	(ii) When $|F'(x)|$ has at most linear growth, the mean-value theorem gives
	\begin{align*}
		\bigg|
		F\Big(x\Big(\frac{j}{N}\Big)\Big) - F\Big(x\Big(\frac{j+\tau}{N}\Big)\Big)
		\bigg| & \Le \Big|x\Big(\frac{j}{N}\Big) - x\Big(\frac{j+\tau}{N}\Big)\Big|
		\bigg(\Big|x\Big(\frac{j}{N}\Big)\Big| +
		\Big|x\Big(\frac{j+\tau}{N}\Big)\Big| + 1
		\bigg),
	\end{align*}
	and Cauchy's inequality gives
	\begin{align*}
		& \mathbb E_{\sigma[x_0]} \Bigg[\bigg|
		F\Big(x\Big(\frac{j}{N}\Big)\Big) - F\Big(x\Big(\frac{j+\tau}{N}\Big)\Big)
		\bigg|\Bigg] \\
		& \quad \Le C_{\beta,m}\sqrt{
			\mathbb E_{\sigma[x_0]}\bigg[\Big|x\Big(\frac{j}{N}\Big) - x\Big(\frac{j+\tau}{N}\Big)\Big|^2\bigg]
			\mathbb E_{\sigma[x_0]} \bigg[\bigg(\Big|x\Big(\frac{j}{N}\Big)\Big| +
			\Big|x\Big(\frac{j+\tau}{N}\Big)\Big| + 1
			\bigg)^2\bigg]
		} \\
		& \quad \Le C_{\beta,m}\sqrt{\frac1N \cdot (|x_0|+1)^2} \Le \frac{C_{\beta,m}(|x_0|+1)}{\sqrt{N}},
	\end{align*}
	which establishes part (ii).
\end{proof}

\subsection{Proof of convergence}
In the periodic case, \eqref{WW bound} and part~(i) of \Cref{lemma: J bound} directly yield
\begin{equation*}
	|\mathcal W_N - \mathcal W| \Le \frac{C_{\beta,m}}{\sqrt{N}},
\end{equation*}
which is the first part of \Cref{lemma: consistency}.

In the confining case, we apply the domain truncation technique from \Cref{section: truncation}. Recall the truncated trace quantity
\begin{equation*}
	\mathcal W_N(R) =
	\sqrt{\frac{m}{2\pi\beta}}\int_{-R}^R \d x_0
	\times \mathbb E_{\sigma_N[x_0]}
	\bigg[
	\frac1N\sum_{j=0}^{N-1} O(x_j) \times
	\exp\bigg(-\frac{\beta}{N}
	\sum_{j=0}^{N-1} V(x_j)\bigg)
	\bigg],
\end{equation*}
together with the truncation error from \Cref{lemma: Z error},
\begin{equation}
	|\mathcal W_N(R) - \mathcal W_N| \Le \frac{C_{\beta,m}}{R}\, e^{-c_{\beta,m}R^2},
	\label{WW R bound}
\end{equation}
in which $C_{\beta,m}, c_{\beta,m} > 0$ are independent of $R$ and $N$.

We now derive the continuum counterpart of \eqref{WW R bound}. Define
\begin{equation*}
	\mathcal W(R) =
	\sqrt{\frac{m}{2\pi\beta}}\int_{-R}^R \d x_0
	\times \mathbb E_{\sigma[x_0]}
	\bigg[
	\int_0^1 O(x(\tau))\d\tau \times
	\exp\bigg(-\beta \int_0^1 V(x(\tau))\d\tau\bigg)
	\bigg].
\end{equation*}
Repeating in the continuous setting the argument that produced \eqref{WRW difference N} gives
\begin{align}
	|\mathcal W(R) - \mathcal W| & \Le
	C_{\beta,m} \int_{[-R,R]^c} \d x_0 \times
	\mathbb E_{\sigma[x_0]}\bigg[\exp\bigg(-
	\beta \int_0^1 V(x(\tau)) \d\tau
	\bigg)\bigg] \notag \\
	& \Le
	C_{\beta,m} \int_{[-R,R]^c} \d x_0 \times
	\mathbb E_{\sigma[x_0]}\bigg[\exp\bigg(-
	\frac{a\beta}{2} \int_0^1 |x(\tau)|^2 \d\tau
	\bigg)\bigg] \notag \\
	& \Le C_{\beta,m}\int_R^\infty \d x_0 \times \int_0^1
	\mathbb E_{\sigma[x_0]} \Big[
	e^{-\frac{a\beta}{2}|x(\tau)|^2}
	\Big] \d\tau \notag \\
	& \Le C_{\beta,m}\int_R^\infty \exp\bigg(-\frac{a\beta x_0^2}{2(1+a\beta^2/(4m))}\bigg) \d x_0 \Le \frac{C_{\beta,m}}{R}\, e^{-c_{\beta,m}R^2}.
	\label{W R bound}
\end{align}

Combining \eqref{WW R bound}, \eqref{W R bound}, and part~(ii) of \Cref{lemma: J bound} via the triangle inequality,
\begin{align*}
	|\mathcal W_N - \mathcal W| & \Le
	|\mathcal W_N - \mathcal W_N(R)| +
	|\mathcal W - \mathcal W(R)| +
	|\mathcal W_N(R) - \mathcal W(R)| \\
	& \Le \frac{C_{\beta,m}}{R}\, e^{-c_{\beta,m}R^2} + \frac{C_{\beta,m}}{\sqrt{N}}\int_{-R}^R (|x_0|+1)\d x_0 \\
	& \Le \frac{C_{\beta,m}}{R}\, e^{-c_{\beta,m}R^2} +  \frac{C_{\beta,m}(R^2+R)}{\sqrt{N}}.
\end{align*}
The choice $R^2 = \frac{1}{2c_{\beta,m}}(\log N+1)$ balances the two terms and yields
$$
|\mathcal W_N - \mathcal W| \Le \frac{C_{\beta,m}(\log N+1)}{\sqrt{N}},
$$
which proves the second part of \Cref{lemma: consistency}.

\section{Alternate operator splitting}
\label{appendix: alternate}

An alternative approach repartitions the Hamiltonian to accommodate confining potentials and achieves the optimal $\mathcal O(1/N^2)$ convergence rate under stronger regularity assumptions. For any constant $a>0$, split $\hat H = \hat H_a + \hat V_a$ by defining
\begin{equation*}
	\hat H_a = \hat H_0 + \frac{a}{2} \hat x^2,
	\qquad 
	\hat V_a = V(\hat x) - \frac{a}{2} \hat x^2.
\end{equation*}
Here, $\hat H_a$ is the Hamiltonian for a quantum harmonic oscillator, for which the partition function $\tr\big[e^{-\beta\hat H_a}\big]$ is finite and analytically known. The remaining term, $\hat V_a$, serves as a modified potential. The Lie--Trotter approximation based on this splitting is
\begin{equation*}
	\mathcal Z_N^a = \tr\Big[
	\big(e^{-\frac{\beta}{N}\hat H_a}
	e^{-\frac{\beta}{N}\hat V_a}\big)^N
	\Big].
\end{equation*}
A path integral representation for $\mathcal Z_N^a$ is derived analogously to the standard splitting, yielding an expectation over the quantum harmonic oscillator:
\begin{equation*}
	\mathcal Z_N^a =
	\tr\big[e^{-\beta\hat H_a}\big] \cdot 
	\mathbb E_{\mu_N^a}
	\bigg[
	\exp\bigg(-\frac{\beta}{N}
	\sum_{j=0}^{N-1} V_a(x_j)\bigg)
	\bigg],
\end{equation*}
where the expectation is taken with respect to the probability distribution
\begin{equation*}
	\mu_N^a(x) \propto \prod_{j=0}^{N-1} \mel{x_{j+1}}{e^{-\frac{\beta}N \hat H_a}}{x_j}.
\end{equation*}
Since $\hat H_a$ describes a quantum harmonic oscillator, the propagator $\mel{x_{j+1}}{e^{-\frac{\beta}{N}\hat H_a}}{x_j}$ is the kernel of a Gaussian process (an Ornstein--Uhlenbeck bridge), and the resulting path measure $\mu_N^a$ is a multivariate Gaussian distribution.

Following the Araki--Lieb--Thirring inequality in \eqref{ALT N}, the successive difference can be expressed as an expectation over paths of length $2N$:
\begin{equation*}
	\mathcal Z_N^a - \mathcal Z_{2N}^a = \frac12\tr\big[e^{-\beta\hat H_a}\big] \cdot
	\mathbb E_{\mu_{2N}^a}
	\bigg[\Big(e^{-\frac{\beta}{2N} \sum_{j=0}^{N-1} V_a(x_{2j})} - 
	e^{-\frac{\beta}{2N} \sum_{j=0}^{N-1} V_a(x_{2j+1})}\Big)^2\bigg] \Ge 0.
\end{equation*}
The successive difference is therefore non-negative and is controlled by the variance of $V_a(x)$ along the path. If the derivatives of $V_a(x)$ are uniformly bounded, the argument from \Cref{theorem: periodic} carries over and yields the optimal $\mathcal O(1/N^2)$ rate. We formalize the regularity condition as follows.

\begin{assumption}[alternate]
	\label{assumption: alternate}
	The modified potential $V_a(x) = V(x) - ax^2/2$ is twice differentiable, and its derivatives are uniformly bounded, i.e.,
	\begin{equation*}
		\sup_{x\in\mathbb R}\Big(|V_a(x)|+|V_a'(x)| + |V_a''(x)|\Big) < +\infty.
	\end{equation*}
\end{assumption}

Under this condition, the periodic-case analysis adapts to the Ornstein--Uhlenbeck bridge and yields an $\mathcal O(1/N^2)$ error bound. We state the result without proof.

\begin{theorem}[alternate]
	Under Assumption~\ref{assumption: alternate}, the Lie--Trotter approximation based on the harmonic oscillator splitting converges with an optimal rate:
	\begin{equation*}
		0 \Le \mathcal Z_N^a - \mathcal Z
		\Le \frac{C_{\beta,m}}{N^2},
	\end{equation*}
	where $C_{\beta,m}>0$ depends on $\beta$, $m$, and the constants in Assumption~\ref{assumption: alternate}, but not on $N$.
\end{theorem}

\bibliographystyle{plain}
\bibliography{references}

@misc{tao2010golden,
	author       = {Tao, Terence},
	title        = {The Golden-Thompson inequality},
	howpublished = {WordPress blog post},
	month        = {July},
	year         = {2010},
	day          = {15},
	url          = {https://terrytao.wordpress.com/2010/07/15/the-golden-thompson-inequality/},
	note         = {Accessed: 2025-10-02}
}

@book{davidson2013statistical,
	title={Statistical mechanics},
	author={Davidson, Norman},
	year={2013},
	publisher={Courier Corporation}
}

@book{tuckerman2023statistical,
	title={Statistical mechanics: theory and molecular simulation},
	author={Tuckerman, Mark E},
	year={2023},
	publisher={Oxford university press}
}

@book{selsto2024computational,
	title={A computational introduction to quantum physics},
	author={Selst{\o}, S{\o}lve},
	year={2024},
	publisher={Cambridge University Press}
}

@misc{feynman1966quantum,
	title={Quantum mechanics and path integrals},
	author={Feynman, Richard Phillips and Hibbs, Albert R and Weiss, George H},
	year={1966},
	publisher={American Institute of Physics}
}

@book{szabo1996modern,
	title={Modern quantum chemistry: introduction to advanced electronic structure theory},
	author={Szabo, Attila and Ostlund, Neil S},
	year={1996},
	publisher={Courier Corporation}
}

@book{nielsen2010quantum,
	title={Quantum computation and quantum information},
	author={Nielsen, Michael A and Chuang, Isaac L},
	year={2010},
	publisher={Cambridge university press}
}

@book{simon2005functional,
	title={Functional integration and quantum physics},
	author={Simon, Barry},
	number={351},
	year={2005},
	publisher={American Mathematical Soc.}
}

@article{suzuki1976generalized,
	title={Generalized Trotter's formula and systematic approximants of exponential operators and inner derivations with applications to many-body problems},
	author={Suzuki, Masuo},
	journal={Communications in Mathematical Physics},
	volume={51},
	number={2},
	pages={183--190},
	year={1976},
	publisher={Springer}
}

@article{ceperley1995path,
	title={Path integrals in the theory of condensed helium},
	author={Ceperley, David M},
	journal={Reviews of Modern Physics},
	volume={67},
	number={2},
	pages={279},
	year={1995},
	publisher={APS}
}

@article{marx1996ab,
	title={Ab initio path integral molecular dynamics: Basic ideas},
	author={Marx, Dominik and Parrinello, Michele},
	journal={The Journal of chemical physics},
	volume={104},
	number={11},
	pages={4077--4082},
	year={1996},
	publisher={American Institute of Physics}
}

@article{ceriotti2010efficient,
	title={Efficient stochastic thermostatting of path integral molecular dynamics},
	author={Ceriotti, Michele and Parrinello, Michele and Markland, Thomas E and Manolopoulos, David E},
	journal={The Journal of chemical physics},
	volume={133},
	number={12},
	year={2010},
	publisher={AIP Publishing}
}

@article{ye2021efficient,
	title={Efficient sampling of thermal averages of interacting quantum particle systems with random batches},
	author={Ye, Xuda and Zhou, Zhennan},
	journal={The Journal of Chemical Physics},
	volume={154},
	number={20},
	year={2021},
	publisher={AIP Publishing}
}

@article{liu2016simple,
	title={A simple and accurate algorithm for path integral molecular dynamics with the Langevin thermostat},
	author={Liu, Jian and Li, Dezhang and Liu, Xinzijian},
	journal={The Journal of chemical physics},
	volume={145},
	number={2},
	year={2016},
	publisher={AIP Publishing}
}

@article{forrester2014golden,
	title={The Golden--Thompson inequality: Historical aspects and random matrix applications},
	author={Forrester, Peter J and Thompson, Colin J},
	journal={Journal of Mathematical Physics},
	volume={55},
	number={2},
	year={2014},
	publisher={AIP Publishing}
}

@article{aujla2003weak,
	title={Weak majorization inequalities and convex functions},
	author={Aujla, Jaspal Singh and Silva, Fernando C},
	journal={Linear Algebra and Its Applications},
	volume={369},
	pages={217--233},
	year={2003},
	publisher={Elsevier}
}

@book{hiai2014introduction,
	title={Introduction to matrix analysis and applications},
	author={Hiai, Fumio and Petz, D{\'e}nes},
	year={2014},
	publisher={Springer Science \& Business Media}
}

@article{neidhardt1998error,
	title={On error estimates for the Trotter--Kato product formula},
	author={Neidhardt, Hagen and Zagrebnov, Valentin A},
	journal={Letters in Mathematical Physics},
	volume={44},
	number={3},
	pages={169--186},
	year={1998},
	publisher={Springer}
}

@article{rogava1993error,
	title={Error bounds for Trotter-type formulas for self-adjoint operators},
	author={Rogava, Dzh L},
	journal={Functional Analysis and Its Applications},
	volume={27},
	number={3},
	pages={217--219},
	year={1993},
	publisher={Springer}
}

@article{doumeki1998error,
	title={Error bounds on exponential product formulas for Schr{\"o}dinger operators},
	author={Doumeki, Atsushi and Ichinose, Takashi and Tamura, Hideo},
	journal={Journal of the Mathematical Society of Japan},
	volume={50},
	number={2},
	pages={359--377},
	year={1998},
	publisher={The Mathematical Society of Japan}
}

@article{takanobu1997error,
	title={On the error estimate of the integral kernel for the Trotter product formula for Schr{\"o}dinger operators},
	author={Takanobu, Satoshi},
	journal={The Annals of Probability},
	volume={25},
	number={4},
	pages={1895--1952},
	year={1997},
	publisher={Institute of Mathematical Statistics}
}

@article{ichinose1997estimate,
	title={Estimate of the difference between the Kac operator and the Schr{\"o}dinger semigroup},
	author={Ichinose, Takashi and Takanobu, Satoshi},
	journal={Communications in mathematical physics},
	volume={186},
	number={1},
	pages={167--197},
	year={1997},
	publisher={Springer}
}

@article{childs2021theory,
	title={Theory of Trotter error with commutator scaling},
	author={Childs, Andrew M and Su, Yuan and Tran, Minh C and Wiebe, Nathan and Zhu, Shuchen},
	journal={Physical Review X},
	volume={11},
	number={1},
	pages={011020},
	year={2021},
	publisher={APS}
}

@article{burgarth2024strong,
	title={Strong error bounds for Trotter and strang-splittings and their implications for quantum chemistry},
	author={Burgarth, Daniel and Facchi, Paolo and Hahn, Alexander and Johnsson, Mattias and Yuasa, Kazuya},
	journal={Physical Review Research},
	volume={6},
	number={4},
	pages={043155},
	year={2024},
	publisher={APS}
}

@article{lu2020continuum,
	title={Continuum limit and preconditioned Langevin sampling of the path integral molecular dynamics},
	author={Lu, Jianfeng and Lu, Yulong and Zhou, Zhennan},
	journal={Journal of Computational Physics},
	volume={423},
	pages={109788},
	year={2020},
	publisher={Elsevier}
}

@article{ye2023dimension,
	title={Dimension-free Ergodicity of Path Integral Molecular Dynamics},
	author={Ye, Xuda and Zhou, Zhennan},
	journal={arXiv preprint arXiv:2307.06510},
	year={2023}
}

@article{audenaert2007araki,
  title={On the Araki--Lieb--Thirring inequality},
  author={Audenaert, Koenraad MR},
  journal={arXiv preprint math/0701129},
  year={2007}
}

@article{araki1990inequality,
  title={On an inequality of Lieb and Thirring},
  author={Araki, Huzihiro},
  journal={Letters in Mathematical Physics},
  volume={19},
  number={2},
  pages={167--170},
  year={1990}
}

@article{ichinose2001norm,
	title={The norm convergence of the Trotter--Kato product formula with error bound},
	author={Ichinose, Takashi and Tamura, Hideo},
	journal={Communications in Mathematical Physics},
	volume={217},
	number={3},
	pages={489--502},
	year={2001},
	publisher={Springer}
}

@article{ichinose2001note,
	title={Note on the Paper “The Norm Convergence of the Trotter--Kato Product Formula with Error Bound” by Ichinose and Tamura},
author={Ichinose, Takashi and Tamura, Hideo and Tamura, Hiroshi and Zagrebnov, Valentin A},
journal={Communications in Mathematical Physics},
volume={221},
number={3},
pages={499--510},
year={2001},
publisher={Springer}
}

@article{becker2025convergence,
	title={Convergence Rates for the Trotter Splitting for Unbounded Operators},
	author={Becker, Simon and Galke, Niklas and van Luijk, Lauritz and Salzmann, Robert},
	journal={Foundations of Computational Mathematics},
	pages={1--54},
	year={2025},
	publisher={Springer}
}

@article{suzuki1991general,
	title={General theory of fractal path integrals with applications to many-body theories and statistical physics},
	author={Suzuki, Masuo},
	journal={Journal of Mathematical Physics},
	volume={32},
	number={2},
	pages={400--407},
	year={1991},
	publisher={AIP Publishing}
}

@incollection{hatano2005finding,
	title={Finding exponential product formulas of higher orders},
	author={Hatano, Naomichi and Suzuki, Masuo},
	booktitle={Quantum Annealing and Other Optimization Methods},
	series={Lecture Notes in Physics},
	volume={679},
	pages={37--68},
	year={2005},
	publisher={Springer}
}

@article{jahnke2000error,
	title={Error bounds for exponential operator splittings},
	author={Jahnke, Tobias and Lubich, Christian},
	journal={BIT Numerical Mathematics},
	volume={40},
	number={4},
	pages={735--744},
	year={2000},
	publisher={Springer}
}

@article{descombes2010exact,
	title={An exact local error representation of exponential operator splitting methods for evolutionary problems and applications to linear {Schr\"odinger} equations in the semi-classical regime},
	author={Descombes, St\'ephane and Thalhammer, Mechthild},
	journal={BIT Numerical Mathematics},
	volume={50},
	number={4},
	pages={729--749},
	year={2010},
	publisher={Springer}
}

@book{lubich2008quantum,
	title={From quantum to classical molecular dynamics: reduced models and numerical analysis},
	author={Lubich, Christian},
	series={Zurich Lectures in Advanced Mathematics},
	year={2008},
	publisher={European Mathematical Society}
}
\end{document}